\documentclass{IEEEtran}
\usepackage[short]{optidef}
\usepackage{lipsum}
\usepackage{cite}
\usepackage{filecontents}
\usepackage[final]{graphicx}
\usepackage{amssymb}
\usepackage{amsmath}
\usepackage{enumerate}
\usepackage[normalem]{ulem}
\usepackage{cancel}
\usepackage{caption}
\usepackage{float}
\usepackage{nameref}
\usepackage{xr}
\usepackage{epstopdf}
\usepackage{tikz}
\usetikzlibrary{shapes.arrows}
\usepackage{rotating}
\usepackage{xcolor}
\usepackage{color}
\usepackage{bm}
\usepackage{graphicx,multirow}

\newcommand{\ffunc}{{f}}

\newcommand{\ic}{{i}}

\newcommand{\Qm}{{\bf Q}}
\newcommand{\Rm}{{\bf R}}

\newcommand{\Hm}{{\bf H}}
\newcommand{\Hme}{{H}}

\newcommand{\yv}{{\bf y}}

\newcommand{\Fm}{{\bf F}}

\newcommand{\norm}[1]{\left\lVert#1\right\rVert}

\DeclareMathOperator*{\argmin}{\arg\!\min}
\DeclareMathOperator*{\argmax}{\arg\!\max}
\colorlet{shadecolor}{blue!20}
\usepackage{tcolorbox} \newtcolorbox{mybox}[1]{colback=green!5!white,colframe=blue!75!black,fonttitle=\bfseries,title=#1}
\tcbuselibrary{breakable}

\begin{document}

\title{Signal Processing for Gigabit-Rate Wireline Communications}
\author{ S.~M.~ Zafaruddin,~\IEEEmembership{Member,~IEEE,}
	Itsik Bergel,~\IEEEmembership{Senior Member,~IEEE,}
	Amir Leshem,~\IEEEmembership{Senior Member,~IEEE} 
	\IEEEauthorblockA{Faculty of Engineering, Bar-Ilan University, Ramat Gan 52900, Israel\\             
		Email: smzafar@biu.ac.il, itsik.bergel@biu.ac.il, leshema@biu.ac.il
	}
	
}
\maketitle

\begin{abstract}
	Signal processing played an important role in improving the quality of communications over copper cables in earlier DSL technologies. Even more powerful signal processing techniques are required to enable a gigabit per second data rate in the  upcoming G.fast standard. This new standard is different from its predecessors in many respects. In particular, G.fast will use a significantly higher bandwidth. At such a high bandwidth, crosstalk between different lines in a binder will reach unprecedented levels, which are beyond the capabilities of most efficient techniques for interference mitigation.  
	In this  article, we survey the state of the art and research challenges in the design of signal processing algorithms for the G.fast system, with a focus on novel research approaches and design considerations for efficient interference mitigation in G.fast systems.  We also detail  relevant VDSL techniques and  points out their strengths and  limitations  for the G.fast system.
\end{abstract}

\section{Introduction}
Digital subscriber lines (DSL)  have evolved into a viable technology for  last mile access in telecommunication networks \cite{starr1999book,timmers2013mag}.  This technology leverages the existing infrastructure of telephone lines to provide affordable broadband services when  deployment of  optical networks is unfeasible or costly. Since its inception in the 1980's,  DSL has been considered an interim technology to fill the gap until the advent of all optical access networks. However, we are far away from this horizon and the time of an all optical
access network has not arrived yet.  DSL thus remains  a widely used  broadband access technology and will play a key role in the convergence of mobile and fixed technologies for next generation networks.

The upcoming  G.fast (fast access to subscriber terminals) standard \cite{g9701, oksman2016} promises  to achieve fiber-like data rates  by exploiting  the higher bandwidth of the telephone lines than the previous standards \cite{g993, g9921}.  G.fast is expected to deliver gigabit speeds over  short loop lengths as anticipated by Cioffi et al. \cite{Cioffi_Gigabit_2003} \cite{cioffi07cupon} more than a decade ago. This new standard is very different from its predecessors in many respects. While fiber is deployed more and more into the network, rewiring the houses with fiber is extremely expensive. The G.fast is a fiber to the distribution point (FTTdp) technology taking fiber to a distribution point (DP) very close to the customer premise equipment (CPE). The new G.fast standard changed some very fundamental design choices used in VDSL and earlier standards. Most notably  technologies such as time-division duplexing (TDD) and reverse-power-feeding (RPF) are used, and  discontinuous operation is incorporated for energy efficient transmissions.

The TDD scheme avoids near-end crosstalk (NEXT) and facilitates asymmetry in downstream and upstream data rates more efficiently. It also simplifies channel estimation in downstream transmissions by exploiting channel reciprocity for channel state information (CSI). The RPF technology simplifies the powering of DP using the power from the customer CPE.  This eliminates the need for a power infrastructure at the DP and reduces deployment costs. Discontinuous operation optimizes energy consumption by incorporating low power modes as well as switching off the  circuitry at the DP corresponding to the users in offline mode.  The G.fast system requires efficient signal processing techniques to harness the benefits of these features.

The G.fast also uses considerably wider bandwidth than the VDSL system which was using the spectrum up to $30$ \mbox{MHz}.  The first generation of the G.fast system uses up to $106$ \mbox{MHz} whereas the  next generation goes up to $212$ \mbox{MHz}.
At this higher bandwidth, G.fast systems suffer from significantly stronger crosstalk due to the electromagnetic coupling with different lines. With increasing frequency, the crosstalk coupling between the lines attains the same strength as the direct path and destroys the diagonal dominance of the channel. This strong crosstalk poses challenges that cannot be resolved with existing VDSL technology. Hence, the G.fast system  requires newer and more powerful techniques for interference mitigation, also known as crosstalk cancellation. 

Signal processing techniques have played an important role in improving the quality of communications over copper cables and  hold the key for future services. This article provides an overview of the current state of the art and research challenges in the design of signal processing algorithms for the G.fast system, with a focus on multi-user crosstalk cancellation schemes. 

Multi-user signal processing  which was referred to as vectoring by Ginis and Cioffi in \cite{ginis2002} enabled VDSL systems to exceed $100$ Mbps. This multi-user signal processing takes place at a central point that concentrates copper wire pairs from many users. The processing is on both the signals transmitted from the central point to the end users and on the signal received from the users. 
Thus, the vectored system resembles the multi-user MIMO more than the standard MIMO. However, the vectored DSL system  is different from wireless MIMO  in few important respects.  The DSL channel has long channel coherence and thus CSI can be estimated fairly efficiently for multi-user processing. Since users are connected with a fixed modem,  channel tracking is also not difficult.  Moreover, the telephone channel offers a fairly good channel, and the signal to noise ratio (SNR) can be as high as  $60$ \mbox{dB} at low frequencies. The fact that DSL systems are baseband eliminates the detrimental effect of phase noise, and therefore allows for very high spectral efficiency, that can reach $2^{15}$ QAM.

As the G.fast standard uses much higher frequencies than its predecessor VDSL standard, it can no longer rely on some of the key features of VDSL. One of these features is the well conditioning of the channel matrix.  For  VDSL frequencies, the DSL channel matrix is column-wise diagonal dominance (CWDD) in the upstream and row-wise diagonal dominant (RWDD) in the downstream \cite{ginis2002}. This characteristic is very convenient for FEXT cancellation, and most techniques for multi-user interference mitigation in VDSL rely on the diagonal dominance of the channel matrix.  This diagonal dominance enables efficient implementation of non-linear as well as linear crosstalk cancelers \cite{ginis2002, cendrillon2006near, cendrillon2007}. The linear zero forcing (ZF) precoder for the downstream and   the linear ZF canceler in the upstream are near optimal for the VDSL channel. These schemes are simple and do not necessarily require transmit power optimization. Even the matrix inversion required for  linear ZF crosstalk cancellation can be avoided through power series expansion of the MIMO channel \cite{leshem2007}. Simple least mean square (LMS) based adaptive algorithms are very efficient and converge quite rapidly for the diagonally dominant VDSL channels \cite{louveaux2010,bergel2010}.

At  high frequencies of G.fast the crosstalk is significant and the channel matrices are no longer diagonal dominant. Thus, many VDSL algorithms either fail or converge very slowly. Moreover, the adaptive schemes of the VDSL system are either not suitable or  are no longer applicable  in the TDD based G.fast system.

This paper has a broad scope. We begin with an overview of single user signal processing and explain how this enables DSL modem to overcome channel impairment. Then we provide an overview of the G.fast channel, and explain the various alternative techniques for crosstalk cancellation.
We conclude with some design considerations, which provide an insight into the techniques used.

\section{Wireline DMT Technology}

Like any  communication system, the performance of DSL systems is limited by several types of impairments. In the following, we  discuss the main ones and describe the traditional and novel techniques implemented by DSL systems to address them. We also discuss key distinguished features of the G.fast technology and show how it is different from its predecessor technologies.

\subsection{Thermal  Noise}
All communication systems are inherently limited by thermal noise  caused by the random movement of electrons in the system. This additive noise is generally modeled as a random Gaussian signal  that is  independent of the transmit signal. 
The effect of  thermal noise cannot be completely avoided, and it sets an upper limit on  communication performance, which is known as the channel capacity \cite{shannon1948}. Attaining channel capacity requires the implementation of powerful error correction codes. Legacy DSL systems employ the relatively simple Reed Somolon (RS) codes as an outer code and trellis coded modulation (TCM) as an inner code with an interleaver between them \cite{g993}. The combined  RS+TCM coding scheme was also chosen for the G.fast $106$ \mbox{MHz} standard. Capacity-approaching low-density parity-check  (LDPC) codes have also been proposed for the G.fast.
However, the design and implementation of a LDPC code with flexible coding and modulation that can operate at the G.fast data rates is an open research challenge. However, in most scenarios, thermal noise is not the main limiting factor, and other impairments must be considered.
\begin{figure}[t]
	\begin{center}
		\includegraphics[width=\columnwidth]{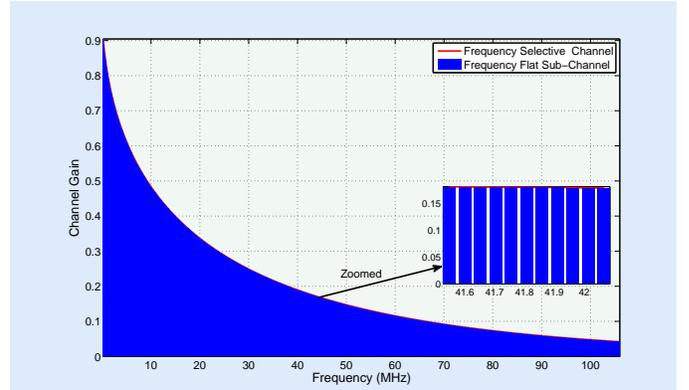}
	\end{center}
	\caption{ Frequency selectivity of the DSL channel is illustrated using the channel gain of a $100$ \mbox{m} CAD55 cable simulated for the G.fast $106a$ profile. The figure shows how the use of $2048$ narrow-band tones of $51.75$ \mbox{KHz} result in an almost frequency flat channel at each tone.}
	\label{DMT}
\end{figure}

\subsection{Inter-Symbol Interference and DMT Modulation}

Over  wide-bandwidths, the telephone channel is frequency selective and exhibits frequency dependent attenuation and delay.  This causes severe inter-symbol interference (ISI) where the communication symbols are prolonged and overlap each other.  To overcome this problem, the recent xDSL standards use a discrete multi-tone (DMT) technique which divides the transmission frequency band into smaller sub-carriers
(also known as tones or frequency bins) \cite{bingham1990multicarrier}. As a result, DMT effectively transforms a broadband frequency selective channel into many frequency-flat narrow band channels as shown in Fig. \ref{DMT}. 

The G.fast has a wider tone width (exactly $12$ times wider than the ADSL/VDSL)  to cover higher bandwidth without increasing the  number of tones. The ADSL system has  $K=256$ tones (over $2.2$ \mbox{MHz}), VDSL (over $30$ \mbox{MHz}) contains $K=4096$ tones, the G.fast $106$ \mbox{MHz} has  $K=2048$ tones, and the G.fast $212$ \mbox{MHz} profile increases the number of tones to $K=4096$.  The DMT symbol also contains a cyclic prefix (CP) that allows a tone separation of $1$ symbol duration without any interference between tones, as long as the CP is longer than the channel memory (see for example \cite{bingham1990multicarrier}). The G.fast has a typical CP length of $L = 320$ samples. 
\begin{figure}[t]
	\begin{center}
		\includegraphics[width=\columnwidth]{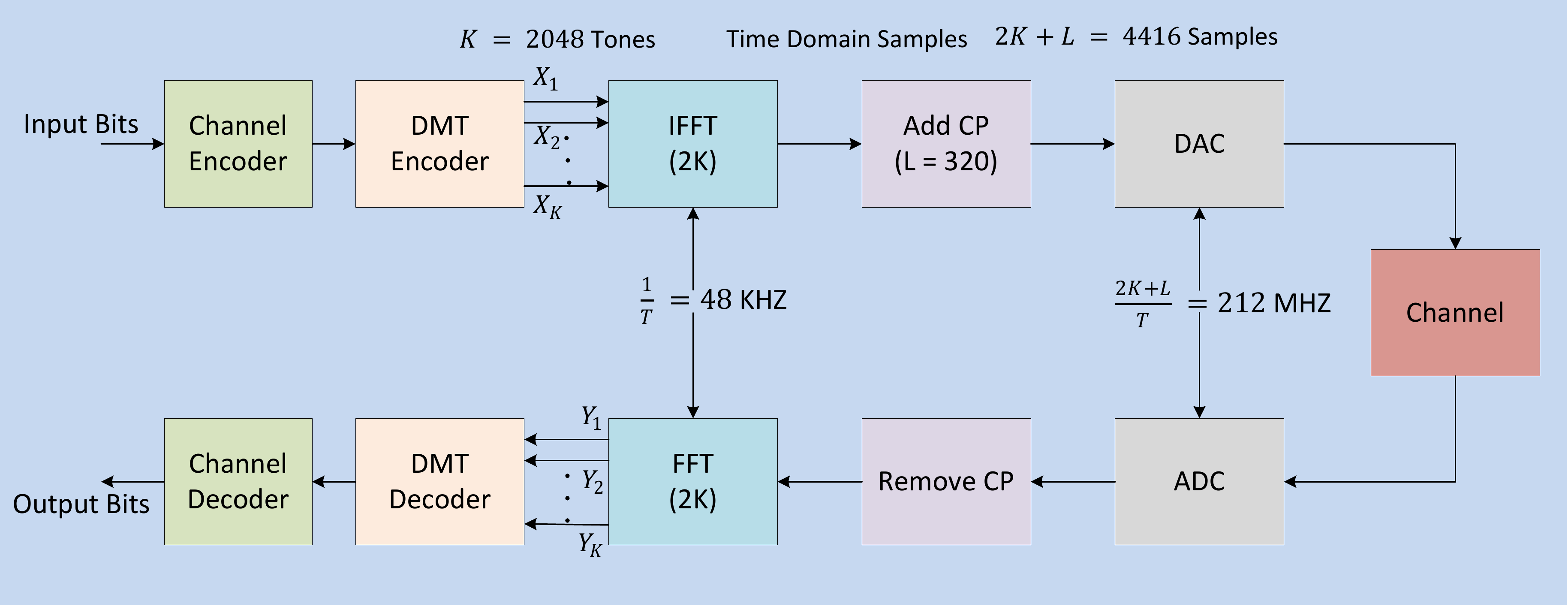}
	\end{center}
	\caption{Discrete multi-tone (DMT) modulation and system parameters for G.fast system. }
	\label{dmt_model2}
\end{figure}

The G.fast has a higher sampling frequency than the VDSL system which operates at  $70$ \mbox{Msps} (million samples per second). The sampling frequency of the G.fast  $106$ \mbox{MHz} profile is $212$ \mbox{Msps}. To convert  $K$ data symbols into a real signal, $2K$ point IFFT is used. The resulting bandwidth is $106$ \mbox{MHz} and the tone-width is $51.75$. Considering the CP, the resulting symbol rate is $48$ \mbox{KHz}  (exactly 12 times faster than ADSL/VDSL). The G.fast $212$ \mbox{MHz} system  has the same symbol rate  but a much higher sampling frequency rate (more than $400$ \mbox{Msps}). A typical  DMT block diagram with G.fast parameters is represented in  Fig.~\ref{dmt_model2}. 

Another important advantage of the DMT is the ability to use a different modulation at each tone. Thus, tones with low SNR will use small constellations, such as QPSK with 2 bits per tone. Tones with high SNRs will use richer  modulations with up to $12$ bits per tone (i.e., $2^{12}=4096$  points). The DMT technique allows the transmission of data symbols without any ISI and without interference between the different tones. Thus, it enables independent processing of each tone. Without loss of generality, in the following we focus on a single tone and  address the other factor that limit system performance.

\subsection{ Near-End Crosstalk (NEXT) and Duplexing Methods }

\begin{figure}[t]
	\begin{center}
		\includegraphics[width=\columnwidth]{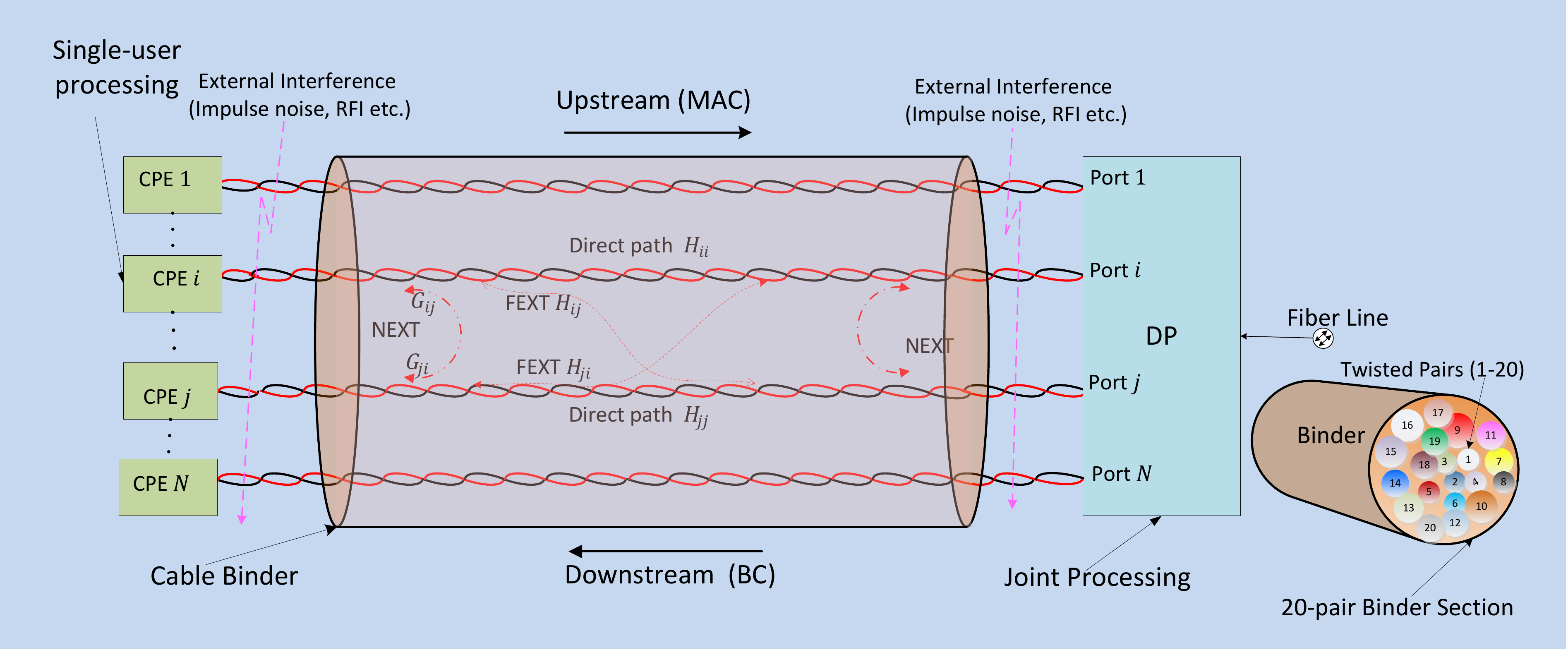}
	\end{center}
	\caption{FEXT and NEXT in a multi-pair DSL binder. The cross section of the cable binder is also shown to demonstrate how  the twisted pairs are  enclosed in the binder. In the upstream, the DSL binder acts as a multiple acces channel (MAC), whereas  broadcast channel (BC) in the downstream. Note that the CPEs are generally situated at different lengths from the DP.}
	\label{crosstalk2}
\end{figure}

\begin{figure}[t]
	\begin{center}
		\includegraphics[width=\columnwidth]{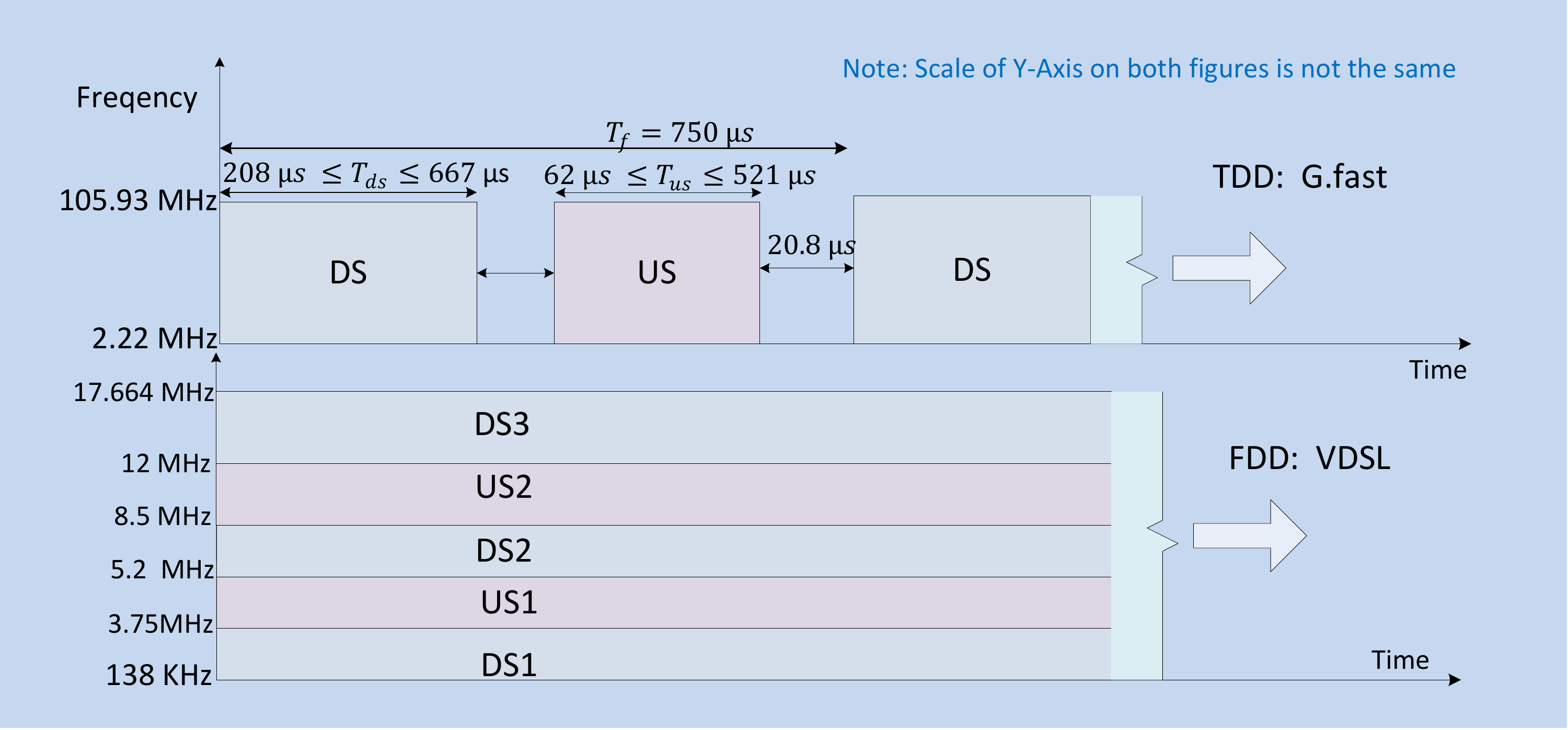}
	\end{center}
	\caption{Duplexing methods for G.fast (TDD system, $106a$ profile)  and  (FDD, $17a$ profile) for VDSL that separate upstream and downstream transmissions to mitigate NEXT in a DSL system.  A typical TDD frame duration is $750$ $\mu s$ comprising $36$ symbols. The number of symbols for downstream ranges from $10$ to $32$ and   $25$ to $3$ symbols for upstream with a single guard symbol between them. }
	\label{tdd}
\end{figure}
A telephone line is composed of two copper lines that are twisted around each other. This twisting reduces the electromagnetic leakage between lines.  However, this is not sufficient and all DSL systems have to cope with electromagnetic coupling signals that increase continuously with frequency.
This, together with large number of closely packed lines in a typical binder lead to
large electromagnetic couplings, which cause significant ``crosstalk". Depending
upon the position of disturbers with respect to the victim receiver, this crosstalk
is classified as  far-end crosstalk (FEXT) or near-end crosstalk (NEXT), as shown in Fig. \ref{crosstalk2}. Near-end crosstalk (NEXT) refers to coupled signals that originate from the same end as the affected receiver.  Hence,  NEXT is interference between upstream signals and downstream signals from different pairs.   As NEXT occurs over short loops, its effect on the receivers is significant. NEXT can be mitigated by using an echo canceler but this is considered impractical in a DSL setup. Thus, all DSL systems separate the upstream and the downstream to avoid the NEXT and echo signals.
The older standards e.g., ADSL and VDSL, separate the upstream and the downstream in the frequency domain, known as the frequency domain duplex (FDD) \cite{g993,g9921}. The latest G.fast standard that further extends the copper bandwidth into hundreds of MHz employs a time-division duplexing (TDD)  where the upstream and downstream are transmitted at different times. The TDD scheme  has access to the full operating spectrum by toggling transmission directions over a time interval, which occurs rapidly and is not visible to the user. It enables dynamic allocation of US and DS resources to efficiently support asymmetric data rates.  This facilitates discontinuous operation and allows for an efficient trade off between throughput and power consumption. The TDD scheme also ensures channel reciprocity for better service provisioning and facilitates channel estimation in downstream.
It also enables a  simplified  analog front-end (AFE) architecture and increases the efficiency of
transmissions.  However, successful implementation of TDD needs a very precise timing and synchronization system at both the transmitter and receiver to avoid interference between two directions. This requires all the nearby modems to be synchronized, a feasible situation at the DP but generally not at the customer premises. The TDD and FDD duplexing systems are illustrated in Fig.~\ref{tdd}.

\subsection{Far-End Crosstalk and Vectored Processing}

\begin{figure}[t]
	\begin{center}
		\includegraphics[width=\columnwidth]{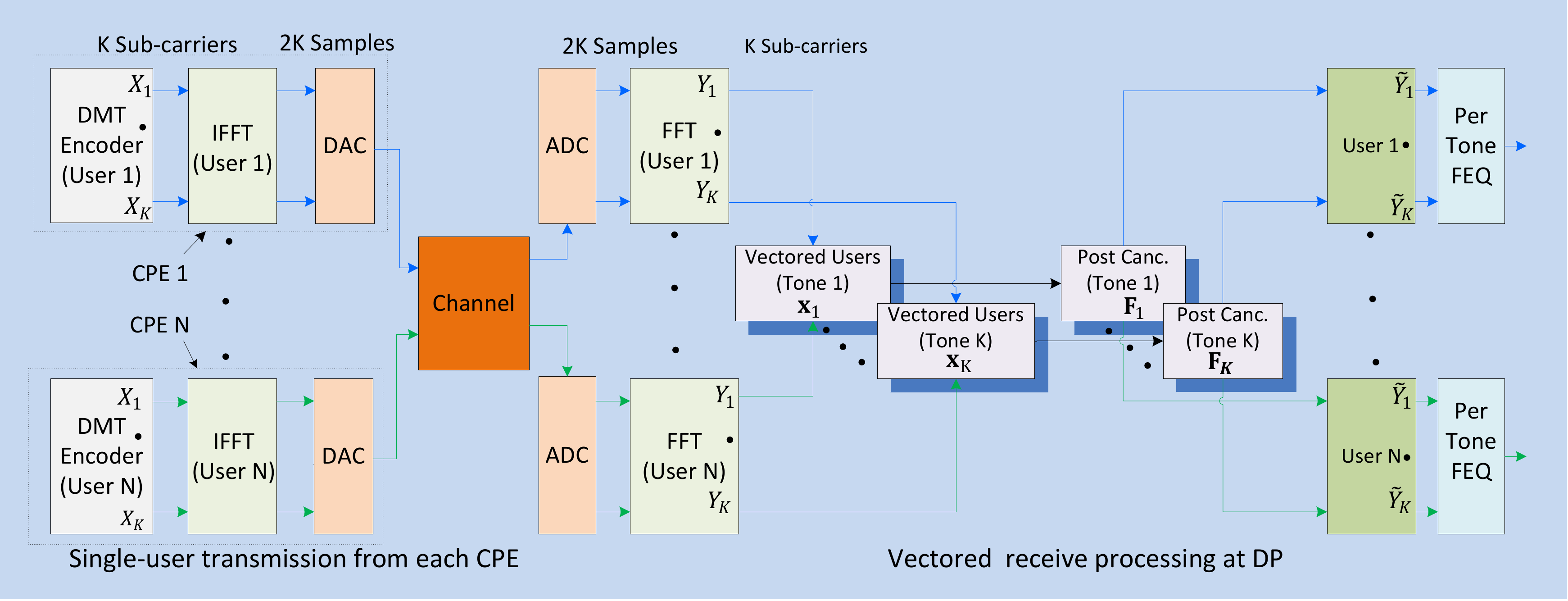}
	\end{center}
	\caption{ Schematic diagram of vectored receive processing in the upstream. There is no signal coordination among CPEs for joint transmit processing. At the DP,  received signals from CPEs  at each tone are collected as a single vector on which a canceler is applied.}
	\label{vectored_receiver}
\end{figure}

\begin{figure}[t]
	\begin{center}
		\includegraphics[width=\columnwidth]{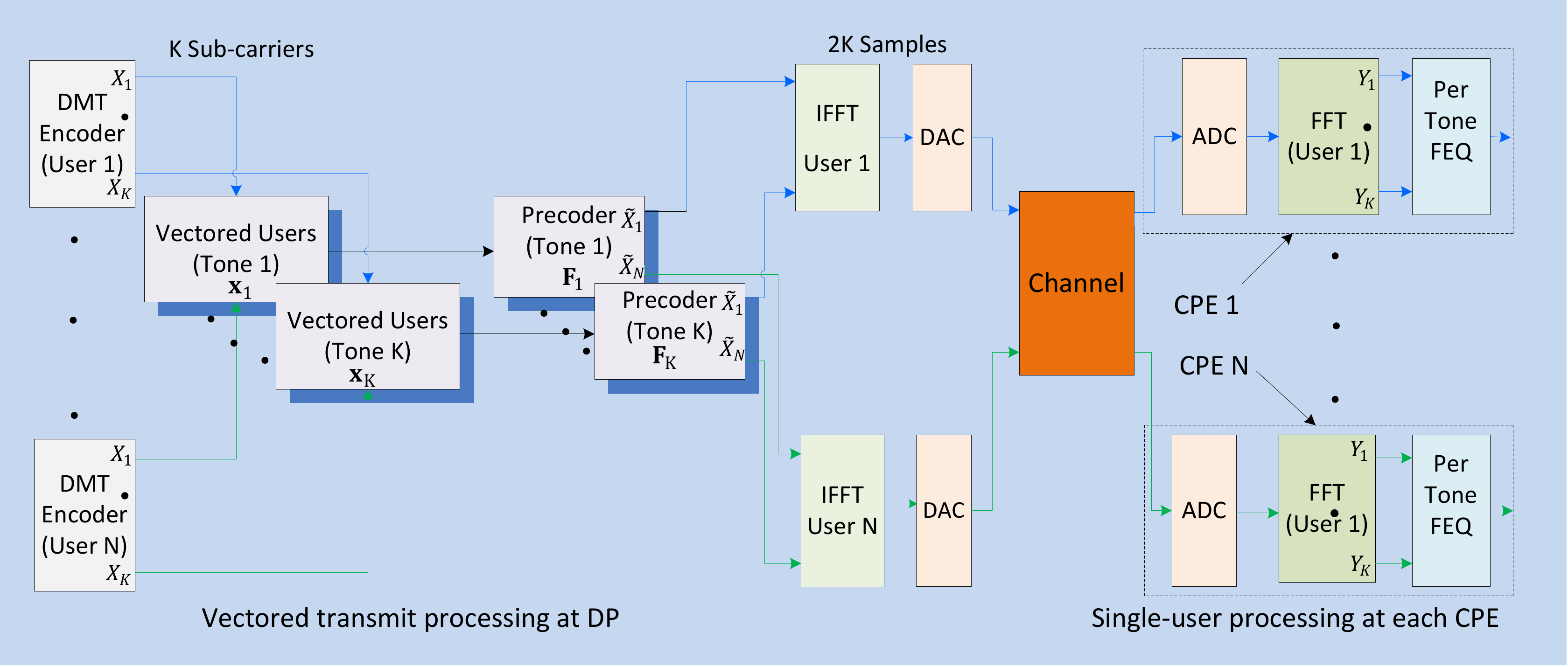}
	\end{center}
	\caption{ Schematic diagram of vectored transmit processing in the downstream.  At the DP,  transmit signals for each user  at each tone are collected as a single vector and a precoder  is applied before transmission.  There is no signal coordination among CPEs for joint receiver processing.}
	\label{vectored_precoder}
\end{figure}
The term FEXT  refers the coupled signals that originate from the end opposite to that of
the affected receiver. FEXT is  thus the interference between upstream signals of different pairs or between downstream signals of different pairs \cite{starr1999book}. The average FEXT power from the $j$ line to the $i$th line can be described as 
\begin{eqnarray}
\mathbb{E}[|H_{ij}|^2] = \chi_{\rm fext} f^{2}d_{ij}\mathbb{E}[|H_{jj}|^2]
\label{eq:power_crosstalk}
\end{eqnarray}
where $\chi_{fext}$ is a constant whose value depends on the physical properties of the copper cable, $\mathbb{E}[|H_{jj}|]$ denotes the attenuation of the disturber, $f$ is the central frequency of the considered tone, and $d_{ij}$ is the coupling length between the victim  and distributer. 

The FEXT creates a linear dependence between the different lines in the binder and hence calls for joint (vector) processing. In the upstream, since  modems are co-located at the DP, it is possible to construct a vector that contains all the received symbols over all lines (at a given tone) and then process them together to reduce the effect of FEXT. On the other hand, in the downstream, the transmitters are co-located and it is possible to  pre-equalize signals and reduce  the crosstalk. This coordinated processing of the  signals over all lines referred to as \emph{vectoring} in \cite{ginis2002}  leads to an enormous rate increase in the DSL system. 
A schematic diagram for  vectored reception is shown in  Fig.~\ref{vectored_receiver}, where the $k$-th tone values $Y_{k,i}$ from all $N$ users are collected to form the vector $\mathbf{y}_k =[Y_{k,1},Y_{k,2},\ldots,Y_{k,N}]^T$, which is used for coordinated signal processing on all components at the DP. The result of this processing can be fed to the conventional single user modem for demodulation and detection. In Fig.~\ref{vectored_precoder}, the $k$-th tone values $X_{k,i}$ from all $N$ users were collected to form the vector $\mathbf{x}_k =[X_{k,1},X_{k,2},\ldots,X_{k,N}]^T$, which is used for  coordinated transmit signal processing on all components at the DP.

In the following, we further describe vector processing, and  focus on a single frequency tone at a time. Hence, without any ambiguity, we drop the tone index from all quantities until the Section V. Note that while the processing of all frequency bins is identical, the performance at the different frequency bins can be significantly different. 

The resulting vector channel model in the upstream is:
\begin{align}
\mathbf{y} = \mathbf{H}\mathbf{x}+\mathbf{w}
\label{eq:up_mimo}
\end{align}
where $\mathbf{x}$ is the vector of all transmitted symbols, $\mathbf{H}$ is the $N\times N$ matrix of all (complex) channel gains at the considered frequency tone and $\mathbf{w}$ represents the additional noise and interference vector. Note that the diagonal elements of $\mathbf{H}$ correspond to the direct paths between the CPEs and the CO on, whereas the off-diagonal elements represent
the FEXT.

In the downstream  joint processing is performed before the transmission. Thus, the channel structure remains the same as in (\ref{eq:up_mimo}), but now $\mathbf{x}$ is the vector of symbols for transmission over all lines (after joint processing) and $\mathbf{y}$ is the vector of all samples received by the different CPEs. Note that in TDD, the channel reciprocity dictates that the channel matrix in the downstream is the matrix transpose of the channel matrix in the upstream. But, in most cases, there are differences in amplitude and phase between the transmission and reception circuits and filters at each end.

FEXT is currently the main limiting factor in DSL systems, and plays an even more crucial role in the upcoming G.fast technology. The various methods to cope with FEXT are further described in the following sections, after a discussion on channel model in the next section.

\section{DSL Channel Models}

Signal processing methods for a communication system require an  accurate characterization  of the underlying transmission medium. For DSL technology,  modeling  twisted pair channels at higher frequencies has been an active field of research for more than two decades as  DSL standards have evolved \cite{g9701, g993, g9921}.  The resulting models are based on extensive measurement campaigns carried out by different laboratories and telecommunication companies to derive parametric cable models \cite{lin80a, werner91,bt1996vdsl,etsi1998,chen1998,karipidis2005,karipidis2006cms,van2011, van2012, acatauassu2014tcom}.
\subsection{Models for Direct and Crosstalk Channels}
The loss in  signal strength transmitted over a cable depends on the physics  of the cable which is implicitly captured in the propagation constant $\gamma(f)$  and  the loop length $l$. The cable insertion loss is modeled as $H(f,l)= \mathrm{e}^{-l\gamma(f)}$ and most of the measurement effort has been directed toward better characterization of $\gamma(f)$. 

Various empirical models are available for different cable types over  VDSL frequencies \cite{bt1996vdsl, etsi1998, chen1998,karipidis2005}. However, the extrapolation of these models for   G.fast  frequencies deviate from the actual measurements. Moreover, the G.fast system incorporates different topologies for deployment using other cable types such as CAT5 and CAD55. In this context, the treatment of the direct path (diagonal channel matrix elements) and the crosstalk paths (off-diagonal elements) is quite different.

Parametric models for the direct channel have been developed for a few cable types and made available in the latest  ITU-T recommendation \cite{g9701}. The ITU-T model  has been validated up to hundreds of \mbox{MHz} using results of an extensive cable measurement campaigns on different cable types \cite{BT}. There is an ongoing effort to improve the existing  models and derive models for other cable types (e.g., the  authors in \cite{acatauassu2014tcom} have proposed  a model for the characteristic impedance and propagation constant of the cables using fewer parameters).

As long as the DSL transmission was limited to a point to point technology such as in ADSL and VDSL2,  the standard model for the crosstalk was called the ``$1$\% worst-case model" which  requires that  there could only be  a $1$\%  chance that the total actual FEXT coupling strength in a real bundle was worse than that obtained with  parametric models  \cite{starr1999book}. This model was based on a log-normal model for  single line crosstalk, and evaluated the $1\%$ worst case for the sum of $49$ interferers. Using the parametric model in (\ref{eq:power_crosstalk}), we can see that  crosstalk coupling depends on the frequency, coupling length of the disturber with the victim, and the attenuation of the interfering signal. For VDSL, the coupling was shown to be  proportional to the frequency.

Similar to the direct path, extrapolation of the crosstalk channel models of the VDSL  shows that the FEXT signal was underestimated at the higher frequencies of the G.fast system. Extensive measurement campaigns are in progress for the standardization of the G.fast crosstalk channel model  \cite{van2011}, \cite{van2012}. Recently, an enhanced version of the  1\% worst case deterministic FEXT model has been proposed for G.fast frequencies \cite{Brink2017}.  In these proposals, the crosstalk channels were shown to have a dual slope  at high frequencies, resulting in a higher FEXT signal than those obtained through extrapolation of the  VDSL models. This model has a lower slope FEXT below $75$ \mbox{MHz} and  higher slope FEXT above $75$ \mbox{MHz} to be consistent with the experimental FEXT data.

\begin{figure}[t]	
	\begin{center}
		\includegraphics[width=\columnwidth]{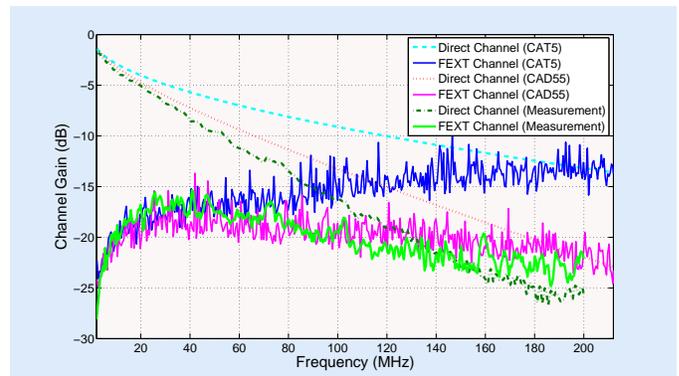}
	\end{center}
	\caption{Direct and crosstalk channel gains for various cables using measurement data \cite{BT} and simulation models. Loop length is  $100$ \mbox{m}. For higher loop lengths, the crosstalk channel gain is higher than the direct channel gain.}
	\label{fig:channel_gain_100m}
\end{figure}

The 1\% modeling is used because the random nature (amplitude variation) of the FEXT channel is much more significant than for the direct channel. However, as G.fast requires more sophisticated crosstalk mitigation, the 1\% modeling is not sufficient and current studies aim to better characterize the randomness of the channel both in terms of amplitude and phase. Various stochastic models of VDSL \cite{sorbara2007, maes09, xuicc09} have adopted the log-normal distribution for the magnitude of the FEXT channel and a uniform distribution for  phase coupling. The authors in  \cite{ginis06asp} also suggested  that the FEXT between different lines in the binder is statistically independent.   Thus, while the modeling of the direct part for G.fast is quite mature, the detailed modeling of the FEXT randomness is currently still under study. All  simulation results presented herein are based on separately modeling  the direct channels \cite{g9701},  FEXT coupling \cite{van2011}, and stochastic parameters \cite{sorbara2007}. 

Fig. 7 presents an example of channel gain and FEXT coupling  of different cables, using stochastic channel models and measurement data for a $0.5$ mm cable with $10$ pairs, measured by BT \cite{BT}.  It can be seen that the CAD55 model has a higher attenuation than the CAT5  model and that the measured cable is closer to the CAD55 model than to the CAT5. Further, the diagonal channel dominates the FEXT channel for VDSL frequencies. However, the FEXT channel becomes very strong at higher frequencies.

\subsection{Diagonal Dominance of Channel Matrix }
\begin{figure}[t]	
	\begin{center}
		\includegraphics[width=\columnwidth]{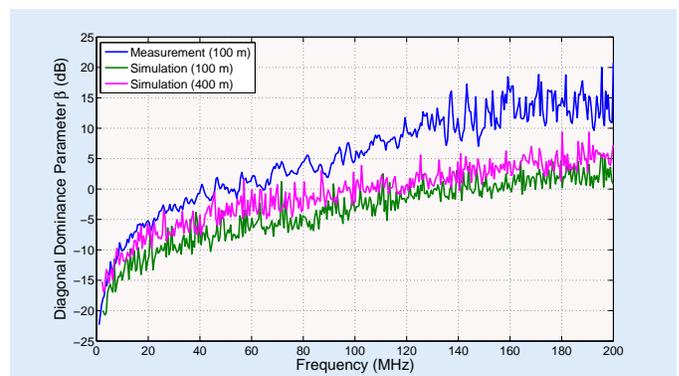}
	\end{center}
	\caption{Diagonal dominance measure $\beta$ as a function of frequency using measurement \cite{BT} and simulation data for a binder with $10$ lines of equal  loop lengths of $100$ \mbox{m} and $400$ \mbox{m}. The VDSL frequencies show strong diagonal dominance whereas the G.fast channel at higher frequency tones is not diagonally dominant.}
	\label{fig:dd_measurement}
\end{figure}

When focusing on a single sub-carrier, the DSL channel matrix, $\mathbf{H}$,  contains the direct channel coefficients for each line (at the diagonal elements) and crosstalk coupling coefficients between each line at  the off-diagonal elements. Each row of the channel matrix represents the crosstalk paths for multiple transmitters for a single receiver whereas each column represents the transmission path  from a single transmitter to the multiple receivers.

Intuitively, one would expect the signal in the desired pair to be much stronger than the signals coupled from other pairs. As shown in Fig. \ref{fig:channel_gain_100m}, this intuition indeed holds true for most VDSL frequencies (up to $30$ MHz),  as the FEXT  among pairs is  insignificant. Thus, for VDSL frequencies, the channel matrix is diagonal dominant, which is very convenient for multi-user communication. In the following we present a short mathematical quantification of the diagonal dominance property. This property will be further discussed in Section \ref{section:crosstalk_techniques} which addresses FEXT cancellation.

We distinguish between two types of diagonal dominant channels. In downstream,  transmitting modems are colocated at the DP and the receiving modems   are situated at different lengths from the DP.  The crosstalk signal from a disturber must propagate through the full length of the victim line to interfere with the victim receiver. Since the insulation between twisted pairs increases the attenuation, each diagonal element dominates its own row, and thus the downstream DSL channel is row-wise 
diagonal dominant (RWDD).  The channel is said to be RWDD if   $\beta_{r}= \max_{i} \sum_{j=1,j\neq i}^N|H_{ij}|/{|H_{ii}|} <1$.  Using the reciprocity principle, the DSL channel in the upstream is column-wise diagonal dominant (CWDD), and is quantified by $\beta_{c}= \max_{i} \sum_{j=1,j\neq i}^N|H_{ji}|/{|H_{ii}|}$.  We further say that the channel is diagonal dominant (DD) if it is both CWDD and RWDD, i.e., if $\beta=\max\{\beta_\mathrm{c},\beta_\mathrm{r}\}<1$.

A diagonal dominant channel ensures  well-conditioned crosstalk channel matrix. Hence, most of the DSL specific research has used this feature in some way or another. Moreover, performance analysis of various algorithms has used the above metrics to bound performance. The diagonal dominance parameter is depicted in Fig.~\ref{fig:dd_measurement} using  measurement data  and simulation results for the CAT5 cable at a length of 100 and 300 meters, for binders with 10 wires. Depending on the scenario, the channels are diagonal dominant ($
\beta$ is less than $0$ dB) up to a frequency which is between $40$ and $100$ MHz. Thus, diagonal dominance holds at these distances for all VDSL frequencies, but does not hold for many G.fast frequencies.

\section{Multi Channel Crosstalk Cancellation Techniques}\label{section:crosstalk_techniques}
\begin{figure}[t]	
	\begin{center}
		\includegraphics[width=\columnwidth]{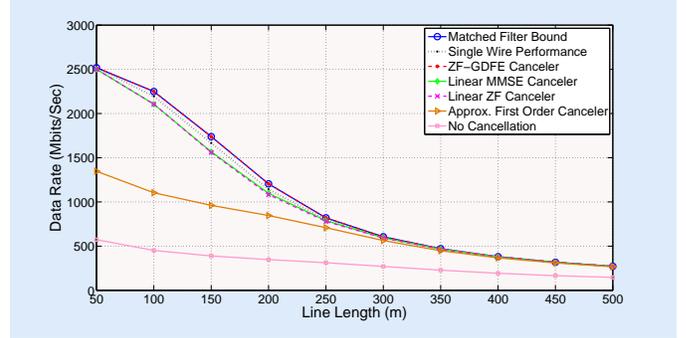}
	\end{center}
	\caption{Average achievable user rate over the whole bandwidth  of $212$ MHz vs. binder length. The binder is composed of $10$ users with equal line lengths..}
	\label{fig:rate_equal_10users_212mhz}
\end{figure}
The various crosstalk cancellation methods available in the literature can be categorized in terms of the coordination among  users in the binder.
If no  coordination is possible,  the binder behaves like an interference channel.  Each receiver decodes its signal independently in the presence of the interference from other users. The advantage of these methods is that they  can  be applied independently on each modem (without any coordination). Unfortunately, these techniques  yield a relatively low data rate for each user in the presence of crosstalk. This is shown in Fig. \ref{fig:rate_equal_10users_212mhz}. It can be seen that  data rate without crosstalk cancellation is just $20\%$ (at $50$\mbox{m}) of what can be achieved with full crosstalk cancellation.

The effect of crosstalk can be reduced using coordinated processing on signals.  In the downstream, the multi-user DSL binder behaves like a broadcast channel (BC) where a single transmitter at the DP generates signals to  geographically dispersed subscribers.  This only enables joint  processing   at the transmitter side. The coordination of received signals is possible in an upstream multiple-access channel (MAC) where a single receiver at the DP receives signals from different users, see Fig. \ref{crosstalk2}. Joint  processing  at both the transmit and receive side of the link requires co-location of both the DP and customer premises modems, which is
possible only in the case of  a bonded DSL system, where a single (typically business) customer uses several twisted pairs to achieve very high rates. In most DSL configurations,  each customer has a single twisted pair, and the customers are situated in different locations. Thus, the DSL system is considered a multi-user MIMO, and all the research on MU-MIMO developed for wireless systems is applicable as well (see for example \cite{Yang2015, Vu2007, Gesbert2007, Rusek2013spm} and references therein).

However, as the rates of DSL systems, and in particular G.fast systems are very high and the number of users can be up to $100$ served simultaneously by the same system, most published algorithms for wireless communication are not feasible, and the DSL community has turned to research low complexity algorithms. With the advent of vectored transmission, there has been a surge of research interest in transceiver design for crosstalk cancellation in DSL systems. Various near-optimal receivers \cite{ginis2002,cendrillon2006near,leshem2007} have been designed to perform crosstalk cancellation with  relatively low complexity.  In the following subsections, we discuss  various  crosstalk cancellation schemes for upstream and  downstream transmission.

\subsection{Crosstalk Cancellation in the Upstream}
Starting with the upstream, we first discuss theoretical performance bounds  and then present various crosstalk cancellation schemes.

\subsubsection{MAC Capacity and Performance Bounds}
The capacity of the MAC channel was derived by Cover decades ago \cite{cover1991}. This capacity is characterized through the achievable rate region, which can be described by a set of $2^N-1$ equations. For example, for the $2$-user case, the rate region is given by:
\begin{align}
\begin{split}
{\cal{R}}_1&<{\cal{R}}_{1}^{\mathrm{mfb}} \\
{\cal{R}}_2&<{\cal{R}}_{2}^{\mathrm{mfb}}\\
{\cal{R}}_1+{\cal{R}}_2&\le \log_2\left(\det(\mathbf{I}_2+\mathbf{H}\mathbf{S}\mathbf{H}^{H}\sigma_w^{-2})\right)
\end{split}
\end{align}
where  $\mathbf{S}$ is a diagonal power matrix,  $R_{1}^{\mathrm{mfb}}$ and $R_{2}^{\mathrm{mfb}}$ are the matched filter bound, given in sub-subsection (ii) below. However, when the number of users is large, and since in practical DSL implementations the coding scheme does not achieve capacity, it is more convenient to use simple performance bounds.  Leshem and Zehavi \cite{leshem2011zehavi} presented  an efficient rate control  for a MAC subject to a PSD mask for the transmitters
in a multi-carrier system, and showed that there is no need for upstream power control as long as the receiver can be kept sufficiently linear.

Although the performance bounds presented below are not achievable, they are quite tight upper bounds in practical DSL scenarios. Hence, they allow us to  quantify the sub-optimality of each algorithm, by evaluating how close it is to the bound. We  demonstrate that (in DSL) better schemes are very close to the bounds, and hence close to optimal. 

\emph{i) Single Wire Performance (SWP)}:  The most intuitive approach is to compare the achievable performance to the case of a single user transmission over a single wire pair. This performance will be denoted hereafter as  single wire performance (SWP).  In this case, the received signal for the $i$-th user (the tested user) is given by: $Y_i= H_{ii}X_i+W_i$, where the single  user is only limited by the additive noise and attenuation of the channel. The additive noise is assumed to be Gaussian distributed. With the assumption of Gaussian distribution on transmit symbols, the Shannon capacity can be derived using the SNR expression.  While the Shannon capacity is achievable, it requires ideal signal processing, and hence is not realizable in practical systems. For this reason, it is customary in DSL systems to model all the system imperfections by a single SNR-gap parameter which is commonly referred as the Shannon gap. The Shannon gap includes all types of imperfections, starting from amplifier noise and ending with  the use of a square QAM constellation instead of the theoretical Gaussian shaping. Thus for a target 
probability of error $10^{-7}$, the SNR gap $\Gamma= 10.75$ \mbox{dB} is taken for DSL systems. For a single tone of width $\Delta_f$,  the SWP of the $i$-th user is given as $\mathcal{R}_i^{\rm swp}=\Delta_f\log_2(1+\Gamma^{-1}P_{x,i}^{}|{H}_i|^2\sigma_{w,i}^{-2})$, where $P_{x,i}$ is the transmit signal power and $\sigma_{w,i}^2$ is the noise variance.

However,  it is worth noting that the  SWP is not an upper bound on  achievable performance. Although FEXT typically degrades  performance,  in some instances the transmission from different modems can be combined coherently through crosstalk between the wires. Hence, the following subsection presents a useful upper bound on the user rate.

\emph{ii) Matched Filter Bound (MFB)}:  The capacity achieved when a single user utilizes both  direct as well as all FEXT coupling for reception  is commonly termed  the matched filter bound (MFB), also known as the single user bound (SUB) in DSL terminology.  As such, all the modems receive from single users, and the received signal for the $i$-th signal becomes:
$\mathbf{y}=\mathbf{h}_iX_i+\mathbf{w}$ where $\mathbf{h}_i$ is the $i$-th column of the channel matrix $\mathbf{H}$.
The optimal processing of the received signal in this case is known as a matched filter (MF) or maximal ratio combining (MRC). This receiver simply requires linear combining of the received signals using:
$\hat{X}_i =\mathbf{h}_i^H\mathbf{y} $
and thus the achievable capacity for the $i$-th user is 
\begin{align}
\mathcal{R}_i^{\rm mfb}= \Delta_f\log_2(1+\Gamma^{-1}\sigma_{w,i}^{-2}P_{x,i}||\mathbf{h}_i||^2)
\end{align}
Note that the power in this bound is the power allowed for a single user and not all the power allowed in the network. This is because this is not a practical scenario, and is simply used to bound the performance for the multi-user case. 

For VDSL system,  there is a marginal difference between MFB and SWP performance since the crosstalk channels  are much smaller than the direct channel gains (see subsection III-B for more details). Since crosstalk increases with frequency, the G.fast already presents a notable gap between the SWP and MFB.
\subsubsection{Non-Linear Crosstalk Cancelers}
In the multi-user case, the MAC capacity can be achieved by detecting a single user at a time and then subtracting the detected signal from the received signal before continuing to detect the next user. This scheme is known as successive interference cancellation (SIC) or generalized DFE (GDFE). It should be noted that  DSL systems have a $6$ \mbox{dB} noise margin, which means that error propagation is unlikely. The optimality of SIC requires linear optimal detection at each stage (using an MMSE receiver) and subtraction of the signal only after  successful decoding of the error correction code as well as proper ordering of the users. Both of these requirements incur significant complexity. In the following we present a simplified version which is near-optimal in DSL systems \cite{ginis2002}.

Consider a decision feedback equalizer (DFE) receiver based on the QR decomposition of the channel $\mathbf{H}$ which can result in rates close to the MFB.  The computation of the QR decomposition of matrix yields $\mathbf{H}= \mathbf{Q}\mathbf{R}$, where $\mathbf{Q}$ is a  unitary matrix and $\mathbf{R}$ is an upper triangular matrix. First, the unitary nature of the matrix $\mathbf{Q}$ is made use of with the linear operation $\mathbf{Q}^H$ on the received vector in (\ref{eq:up_mimo}) to get a rotated version of the received vector 
\begin{align}
\mathbf{z} = \mathbf{Q}^H\mathbf{y} = \mathbf{R}\mathbf{x}+\mathbf{Q}^H\mathbf{w}
\end{align}
As the noise is Gaussian, independent and identically distributed (i.i.d.) between the lines, the multiplication with the  unitary structure of matrix $\mathbf{Q}$ does not change the statistics of the noise. Now, the upper triangular nature of the matrix  $\mathbf{R}$ is made use of, and
the symbols  are estimated starting with the last row. 
The cancellation operation on the $m$-th signal is given as
\begin{align}
\hat{X}_m = Z-\sum_{i=m+1}^{N} \frac{[\mathbf{R}]_{m,i} }{[{\mathbf{R}}]_{m,m}}\hat{X}_i,~~ m = N, N-1 \cdots 1
\label{eq:dfe}
\end{align}
The DFE is sensitive to error propagation: any error in the decision (based on $\hat{X}_m$ in (\ref{eq:dfe})) increases the probability of error for subsequent user detection. However, with proper matching of the modulation and the signal to noise ratio, the error probability can be made small enough so that the error propagation is negligible. In this case, the spectral efficiency for the $i$-th user can be expressed as
\begin{align}
{\cal{R}}_i^{\rm dfe} = \log_2(1+\Gamma^{-1}P_{x,i}|{R}_{ii}|^2\sigma_{w,i}^{-2})
\end{align} Obviously,  user performance is significantly affected by  user ordering (i.e., which user is detected first and which later). For a detailed discussion of various user ordering schemes see \cite{chen2007optimized}.

\subsubsection{Linear Crosstalk Cancelers}
To reduce complexity the straightforward approach is to use linear crosstalk cancelers. The use of linear receivers has attracted  most of the  research interest for FEXT cancellation.  There are several variants of  linear receivers, for example based on the criterion of  zero forcing (ZF) and the minimum mean square error (MMSE). These structures  are much simpler due to the absence of feedback operations.
The zero forcing canceler, as implied by its name, attempts to cancel the self-crosstalk assuming that this is the only disturbance present (thus ignoring even the AWGN and any other kind of interference).  It does this by the application of the inverse operator $\mathbf{F}_{\rm zf} = \mathbf{H}^{-1}$ at the receiver. The application of a linear ZF canceler $\mathbf{H}^{-1}$ on the received signal $\mathbf{y}$ in (\ref{eq:up_mimo}) cancels the crosstalk in the upstream DSL system.  The $i$-th user receives a crosstalk-free signal: $Y_i = X_i+\mathbf{h}_{i}^{\rm inv}\mathbf{w}$, where $\mathbf{h}_{i}^{\rm inv}$ is the $i$-th row of  $\mathbf{H}^{-1}$.  It can be seen that  ZF processing amplifies the power of  additive noise and interference  such that the resultant noise  power becomes $\sum_{j=1}^N \left|[\mathbf{H}^{-1}]_{i,j}\right|^2{\sigma}_{w,i}^2$. Thus, the resulting performance is strongly dependent on the condition number of the channel matrix, and becomes very poor if the matrix is close to singular.

The linear MMSE   canceler $\mathbf{F}$ minimizes the mean square error (MSE) between the output of the canceler and the true value i.e.,  $\argmin_{\mathbf F}  \mathbb{E}[\norm{\mathbf{x}-\mathbf{Fy}}^2]$ is given as \cite{verdu1998}
\begin{align}
\mathbf{F}_{\rm mmse} = (\mathbf{H}^{H}\mathbf{H}+\sigma_{w}^2/P_x\mathbf{I})^{-1}\mathbf{H}^{H}
\end{align}
and results in the spectral efficiency as
\begin{align}
{\cal{R}}^{\rm mmse}_i =\log_2( \frac{\Gamma^{-1}P_x/\sigma_{w}^2}{[(\mathbf{H}^{H}\mathbf{H}+\sigma_{w}^2/P_x\mathbf{I})^{-1} ]_{ii}})
\end{align}

The implementation complexity of MMSE is almost identical to that of ZF, and  is generally better than the ZF solution. The use of MMSE requires the knowledge of the noise covariance matrix. However, any under-estimate of this matrix will still yield  better performance than ZF.

Nevertheless, the advantage of MMSE is negligible for DD channels as well as when the SNR is very high. Thus, the MMSE only performs  significantly  better than the ZF  at high frequency tones.

\begin{tcolorbox}[colback=green!5!white,colframe=blue!75!black,fonttitle=\bfseries,title= A. Linear Cancelers for Upstream DSL System,breakable]	
	\par To illustrate the operation of linear crosstalk cancelers, consider the simple case of $2$ users. 	The users transmit with equal power  $P_x$ over twisted pairs of equal length and experience the equal power of additive noise $P_w$. The direct channel gain for both the users is $H_d$, and the crosstalk coupling is $\alpha H_d$. Thus, in this simple case we have $\beta=\beta_r=\beta_c=\alpha$ and hence, $\alpha$ is a diagonal dominance  parameter. The transmit and received signal vectors can be expressed as:
	\[ \left[ \begin{array}{c}
	Y_1 \\
	Y_2\\
	\end{array} \right]
	= 
	H_d\left[ \begin{array}{cc}
	1 & \alpha \\
	\alpha & 1\\
	\end{array} \right]
	\left[ \begin{array}{c}
	X_1 \\
	X_2 \\
	\end{array} \right]
	+\left[ \begin{array}{c}
	W_1\\
	W_2\\
	\end{array} \right].
	\tag{A1}
	\]
	\hspace{+0.3cm}  Without crosstalk canceling,	the crosstalk signal interferes with the reception of the desired signal, which is now received in the presence of a noise term of power $P_w$ and an interference term of power $\alpha^2H_dP_x$ decreases  the rate. Thus, the spectral efficiency for each is: 
	\begin{align}
	{\cal{R}}^{\rm fext}= \log_2 (1+\frac{\rm{SNR}_{\rm awgn}}{2\alpha^2\rm{SNR}_{\rm awgn}+1})\tag{A2}
	\end{align}
	where  ${\rm{SNR}_{\rm awgn}}= {H_d^2P_x}/{P_w}$. For purposes of  comparison, note that  the single wire performance is	${\cal{R}}^{\rm swp}= \log_2 (1+\rm{SNR}_{\rm awgn})$ whereas the matched filter bound is even higher: $\mathcal{R}^\mathrm{mfb}=\log_2(1+(1+\alpha^2)\rm SNR_{awgn})$. Thus, for high SNR, even small values of $\alpha$ can cause  significant performance degradation. 
	
	\hspace{+0.3cm} 	The ZF crosstalk canceler uses the inverse of the channel  matrix as receiver preprocessing:
	\[ 
	\mathbf{F}_{\rm zf}=\mathbf{H}^{-1}= 
	\frac{1}{H_d(1-\alpha^2)}\left[ \begin{array}{cc}
	1 & -\alpha \\
	-\alpha & 1\\
	\end{array} \right].
	\tag{A3}
	\]
	Thus, the estimate of the symbol of user 1 is: 	
	\begin{align}
	\hat{X}_1 = X_1+\frac{1}{H_d(1-\alpha^2)}W_1-\frac{\alpha}{H_d(1-\alpha^2)}W_2.
	\tag{A4}
	\end{align}
	It can be seen that  ZF processing removes the effect of crosstalk but enhances the  noise by folding noise from other users. The spectral efficiency for the ZF canceler for each user is given as
	\begin{align}
	{\cal{R}}^{\rm zf} = \log_2\left(1+\rm{SNR}_{\rm awgn}\frac{(1-\alpha^2)^2}{1+\alpha^2}\right).
	\label{eq:zfbox}\tag{A5}
	\end{align}
	In contrast to the no cancellation case, the ZF canceler performs very well whenever $\alpha\ll 1$ and approaches  SWP performance. On the other hand, the ZF canceler performs very poorly when the channel matrix is close to singular $\alpha\rightarrow 1$.

	\hspace{+0.3cm} The MMSE canceler for this simple matrix results in the spectral efficiency:
	\begin{align}
	{\cal{R}}^{\rm mmse} = \log_2\left(\rm{SNR}_{\rm awgn}\frac{(1+\alpha^2+\frac{1}{\rm{SNR}_{\rm awgn}})^2-4\alpha^2}{1+\alpha^2+\frac{1}{\rm{SNR}_{\rm awgn}}}\right).
	\label{eq:mmsebox}
	\tag{A6}
	\end{align}
	Comparing 	(\ref{eq:zfbox}) and (\ref{eq:mmsebox}),  the performance of both MMSE and ZF coincides at high SNR. However, if the signal experiences  situations where the attenuation is large or external interference dominates the receiver noise (i.e. lower AWGN SNR), the MMSE canceler performs better than the ZF canceler with a minor increase in complexity.

	\centering
	{\includegraphics[width=\linewidth]{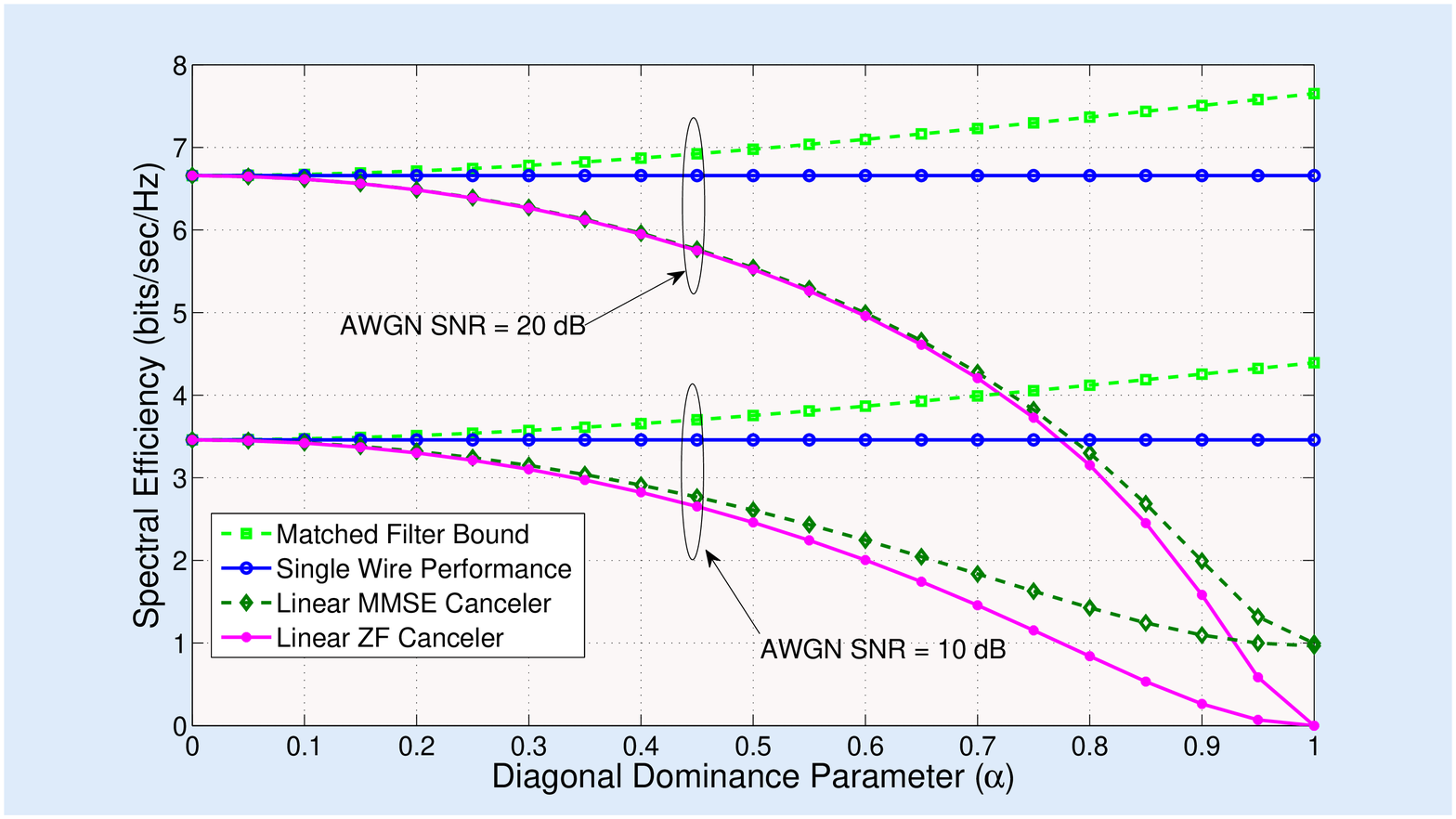}}
	\captionof*{figure}{Fig. S1: Performance of linear cancelers versus the DD parameter ($\alpha$) for two users with  a AWGN SNR of $10$ dB and $20$ dB. For low crosstalk coupling (i.e., $\alpha \ll 1$ ), the MFB is close to the SWP and the SWP metric is sufficient. For large $\alpha$, the MFB becomes more relevant as a metric for performance comparison.  The figure shows that for a SNR of $20$ \mbox{dB} (or higher) the MMSE has practically no advantage over ZF, except for the case  where the channel matrix is very close to singular. For lower SNRs (e.g. $10$ \mbox{dB}) the advantage of the MMSE is greater, and the MMSE canceler performs better than the ZF.}
	
\end{tcolorbox}

\subsection{Crosstalk Cancellation in the Downstream}

In the downstream, crosstalk is pre-compensated for before the transmission of signals. While the downstream processing  attempts to cancel the crosstalk it  also has another consideration  to deal with: the precoded signals at each line should operate within the assigned power spectral mask. Another important difference from the upstream is the lack of channel estimation at the central office.  In this section, we assume a perfect channel for precoding and deal with the cancellation performance of various precoders. Some notes on channel estimation are given in Section V.
\subsubsection{BC Capacity}
The capacity region of the Gaussian BC has been derived by many authors \cite{yu2004, viswanath2003, caire2003, vishwanath2003it, weingarten2006}. Their work was based on the concept of dirty paper coding \cite{cover1991} which makes it possible to transmit data without degradation in the presence of interference that is known to  the transmitter. Equivalently, this capacity can also be achieved using lattice precoding, as discussed by Erez et al. \cite{zamir2002}. 

The derivation of the Gaussian BC capacity is beyond the scope of this review paper. Hence, as in the upstream we will focus on  single wire performance (SWP) and the matched filter bound (MFB).
The single wire performance for the downstream can be similarly derived as in the upstream. The MFB   in downstream transmission is the capacity when all the modems transmit to a single user.  The  MFB for the $i$-th user 
\begin{align}
\mathcal{R}_i^{\rm down}= \Delta_f\log_2(1+\Gamma^{-1}\sigma_{w,i}^{-2}P_{x,i}||\mathbf{h}_i||^2) 
\end{align}where $\mathbf{h}_i$ is the $i$-th row of the downstream channel matrix $\mathbf{H}$, and $P_{x,i}$ is the allowed transmission power for each line (at the considered tone). In the multi-user case the single-user bound can be achieved through non-linear dirty paper coding.
\subsubsection{Non-Linear precoding}
As stated above, optimal non-linear processing can asymptotically achieve the sum rate capacity in the downstream, using dirty paper coding  or multi-dimensional lattice precoding. However, both approaches require  high implementation complexity, and hence are not considered for DSL. A simpler non-linear scheme which is considered for DSL is the Tomlinson-Harashima precoder \cite{ginis2002}, which can be viewed as 1-dimensional lattice precoding. 

Recalling that the FEXT is modeled through the channel matrix, \eqref{eq:up_mimo}, the THP cancels the interference using the the  QR decomposition of the conjugate transpose of the channel matrix, given by  $\Hm^H=\Qm \Rm$, where $\mathbf{Q}$ is an unitary matrix and $\mathbf{R}$ is an upper triangular matrix. The precoding operation is divided into two  parts. The modulated signal ${\mathbf{x}}$ first undergoes a pre-cancelation of the interference using  the elements of $\mathbf{R}$ and a modulo operation, and is then rotated to cancel the channel rotation using the matrix $\mathbf{Q}$.

To remove the interference associated with previous users from the symbol of the $m$-th user, the precoding operation on the $m$-th symbol evaluates:
\begin{align}
\tilde{X}_m = X_m-\sum_{i=1}^{m-1} \frac{[\mathbf{R}^{H}]_{i,m} }{[{\mathbf{R}^{H}}]_{m,m}}{\tilde{X}}_i,~~ m = 1\cdots N .
\label{eq:thp}
\end{align}
Then, to lower the increase in required power, the symbol undergoes a modulo operation:
\begin{align}
{\tilde{X}}_m = X_m \mod 2A 
\end{align}
where the modulo operation is defined such that its result will be within a square with an edge of $2A$ centered at the origin of the complex plane. 
After collecting all symbols, ${\tilde{\mathbf{x}}}=[{\tilde{X}}_1,\ldots,{\tilde{X}}_N]^T$, the resulting vector is rotated by applying $\mathbf{Q}$. Given that the we used the QR decomposition of $\mathbf{H}^H$ and the the matrix $\mathbf{Q}$ is hermitian, the resulting received signal is: $\mathbf{y}=\rm \mathbf{R}^H{\tilde{\mathbf{x}}}+\mathbf{w}$.
By comparing the effect of the channel with the pre-cancellation operation in \eqref{eq:thp}, it can be seen that all interference between users is eliminated. The only step left is to normalize the signal and to reciprocate the modulo operation by an additional modulo operation  at the receivers. Thus, the estimated symbol of the $m$-th user is 
\begin{align}
	\hat{X}_m =\frac{Y_m}{R_{m,m}}\mod 2A= \left[X_m+\frac{W_m}{R_{m,m}}\right]\mod 2A.
\end{align}

Thanks to the modulo operations, the THP can cancel all  the interference at almost no cost. This contrasts with linear precoding schemes (see the next sub-subsection) that can also  remove the interference, but at a power cost that can be significant. The price of the modulo operation comes from the effective change of the channel, so that the Gaussian signaling is no longer optimal. This loss is called the shaping loss, and is at most $1.5$ dB. Moreover, although Gaussian signaling maximizes the achievable rate, all DSL schemes are limited to square QAM modulations, and hence lose most of this shaping loss regardless of the interference cancellation method. Thus, the actual loss of the modulo operation is negligible.

\begin{tcolorbox}[colback=green!5!white,colframe=blue!75!black,fonttitle=\bfseries,title= B.  Tomlinson-Harashima Precoder (THP),breakable]
The THP is an efficient but simple method to remove the interference between the transmitted symbols with almost no cost. While the equations that describe the THP may be somewhat intimidating, the principle of operation is quite simple. In the following we illustrate the operation of the THP for the simple case of  2 users.   
	
	\hspace{+0.3cm}  Consider a downstream transmission for two users. For simplicity, we consider the real valued channel matrix 
	\[      \mathbf{H}=\left[ \begin{array}{cc}
	1 & \alpha \\
	\alpha & 1
	\end{array} \right].
	\tag{B1}
	\]  
	The QR decomposition of the transpose of $\mathbf{H}$ is: 
	\[
	\mathbf{Q}=           \frac{1}{1+\alpha^2}\left[ \begin{array}{cc}
	1 & -\alpha \\
	\alpha & 1
	\end{array} \right]
		\]
		\[
		 \mathbf{R}=\frac{1}{\sqrt{1+\alpha^2}}        \left[ \begin{array}{cc}
	1+\alpha^2 & 2\alpha \\
	0 & 1-\alpha^2
	\end{array} \right]\tag{B2}.
	\]
	\normalsize
	\hspace{+0.3cm}  Let us assume  that the symbols for transmission for the $2$ users are $X_1 $ and $X_2$. Using (\ref{eq:thp}), the symbols are precoded as ${\tilde X}_1=X_1 $ and ${\tilde X}_2 =  X_2- 2\alpha/(1-\alpha^2){X}_1= A+2\alpha/(1-\alpha^2)A$. Ignoring the modulo operation for a while, the symbol vector $\mathbf{\tilde x}=[\tilde X_1,\tilde X_2]$ is rotated by $\mathbf{Q}^T$ and the transmitted vector is $ \mathbf{Q}\mathbf{\tilde x}$. Recalling that the matrix $\mathbf{Q}$ is unitary, we have $\mathbf{Q}\mathbf{Q}^T=\mathbf{I}$, and thus:
	\begin{align}
	\begin{split}
	\mathbf{y}&=\mathbf{H}\mathbf{Q}\mathbf{\tilde{x}}+\mathbf{w} 
	\\
	&	=\mathbf{R}^T\mathbf{Q}^T\mathbf{Q}\mathbf{\tilde{x}}+\mathbf{w}
	\\
	&=\mathbf{R}^T\mathbf{\tilde{x}}+\mathbf{w}
	\end{split}
	\tag{B3}
	\end{align}
	Thus, we get $Y_1=R_{11}X_1+W_1$ and 
	$Y_2 = R_{2,2}X_2+W_2 $. Hence, both symbols can be received by the two separate receivers without any interference. A schematic diagram is presented in Fig. 11. 
	
	\hspace{+0.3cm}  However, in the description above, we ignored the power requirement of this transmission. In particular, assuming that the symbols are statistically independent and with equal power, the transmission of ${\tilde X}_2$ requires a power which is  $1+ 4\alpha^2/(1-\alpha^2)^2$ the power $X_2$. For some values of $\alpha$ this will result in extreme waste of power. 
	
	\hspace{+0.3cm}  To avoid this power loss, we use the modulo operation. Assume that the transmit symbols lie in the range of $[-A, A]$. Thus, instead of using $\tilde X_2$, we use  $\tilde X_2-2An$ where $n$ is an integer  chosen such that the resulting processed symbol is also in the range of $[-A,A]$. At the receiver we again use  the knowledge of the range of transmission values to reconstruct $X_2$. The estimate of $X_2$ is constructed as $\hat X_2=  Y_2/R_{22}+2An$ and $n$ is easily found by requiring that the symbol is  inside the allowed range  of $[-A,A]$. The cost of the modulo operation is slightly lower power efficiency, which is typically negligible compared to the reduction of the interference.

	{\centering
		{\includegraphics[width=\columnwidth]{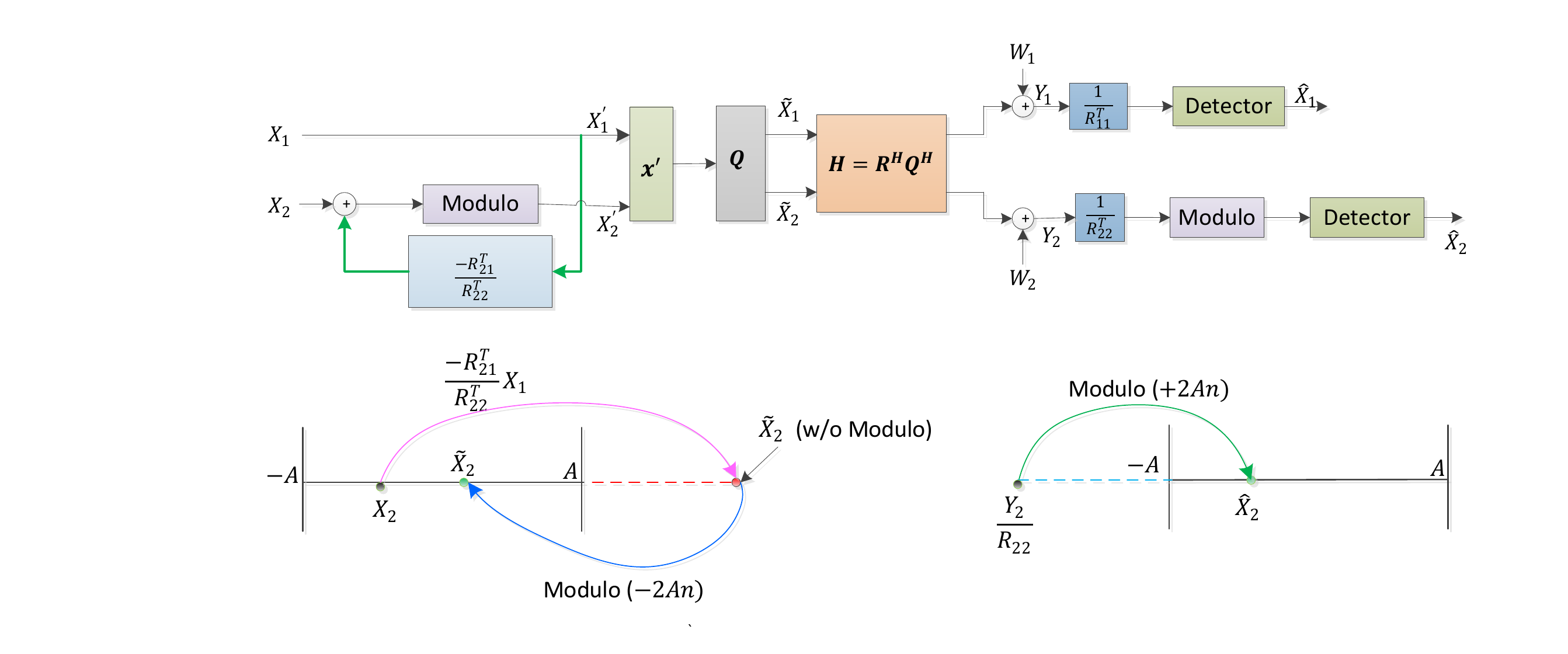}}
		\captionof*{figure}{Fig. S2: Succesive pre-compensation using QR factorization and the Tomlinson-Harashima modulo operation for a simple $2 \times 2$ channel.}}	
	
\end{tcolorbox}

\subsubsection{Linear Precoders}
The equivalent of the MMSE canceller for the downstream is typically termed the diagonal loading precoder or signal to leakage ratio (SLNR) precoder \cite{patchar2012}. As in the upstream, this precoder will typically perform slightly better than the ZF precoder, using almost the same implementation complexity. However, this precoder has not been studied extensively in the context of DSL, and hence will not be discussed here in more detail. 

The linear ZF precoder is a simple technique which pre-compensates the true symbol vector ${\mathbf{x}}$  with  the inverse of the channel matrix, $\mathbf{F}_{\rm zf}=\mathbf{H}^{-1}{\rm diag} (\mathbf{H}) \mathbf{G}$, such that the precoded signal vector becomes $\tilde{\mathbf{x}}=\mathbf{F}_{\rm zf}{\mathbf{x}}$ \cite{cendrillon2007}. 
The  diagonal  scaling matrix $\mathbf{G}$ is chosen such that the total transmit power on each line satisfies the power mask constraint tightly; therefore, each element $G_{ii}$ can be different  \cite{maes2014icc} \cite{neckebroek2015icc}.

\begin{tcolorbox}[colback=green!5!white,colframe=blue!75!black,fonttitle=\bfseries,title= C. Transmit Power Scaling in Linear ZF Precoder,breakable]	
	To illustrate the transmit precoding and necessity of gain scaling,  we consider a downstream transmission of two users with a real-valued channel matrix:
	\[ 
	\mathbf{H}= H_{d}\left[ \begin{array}{ccc}
	1 & \alpha \\
	\alpha & 1\\
	\end{array} \right] \tag{C1}.
	\]
	The ZF linear precoder without gain scaling  $\mathbf{F}_{\rm zf}= \mathbf{H}^{-1}{\rm diag} (\mathbf{H})$:
	\[ 
	\mathbf{F}_{\rm zf}= \frac{1}{1-\alpha^2}\left[ \begin{array}{ccc}
	1 & -\alpha \\
	-\alpha & 1\\
	\end{array} \right].
	\tag{C2}
	\]
	
	The signals after precoding becomes:
	\begin{align}
	\begin{split}
	\tilde{X}_1= \frac{1}{1-\alpha^2} X_1-\frac{\alpha}{1-\alpha^2} X_2.\\
	\tilde{X}_2= \frac{1}{1-\alpha^2} X_2-\frac{\alpha}{1-\alpha^2} X_1.
	\end{split}
	\tag{C3}
	\end{align}
	\hspace{+0.3cm} 	If the users transmit with the maximum power $P^{\rm mask}$, the transmit powers after precoding for both the users are amplified by a factor of $\frac{1+\alpha^2}{(1-\alpha^2)^2}$.
	It is easy to see that the precoded signals violate the PSD mask for $\alpha\neq0$.  Proper scaling of the precoded signals $\tilde{X}_1$ and 	$\tilde{X}_2$ is required  before transmission.
	The elements of diagonal scaling matrix are computed as $G_{11}= G_{22}=1/\argmax_{i}\norm{[\mathbf{H}^{-1}{\rm diag} (\mathbf{H})]_{\rm row~i}}_2= \frac{(1-\alpha^2)} {\sqrt{1+\alpha^2}}$. After scaling the precoded signal, the received signals are free from crosstalk but incur a power penalty: $Y_1 = G_{11}H_dX_1+W_1$ and $Y_2 = G_{22}H_dX_2+W_2$.  The spectral efficiency of the users:
	\begin{align}
	{\cal{R}}_1^{\rm zf}= {\cal{R}}_2^{\rm zf}= \log_2\left(1+\rm{SNR}_{\rm awgn}\frac{(1-\alpha^2)^2}{1+\alpha^2}\right)
	\tag{C4}
	\end{align}
	\hspace{+0.3cm} 	It can be seen that with negligible $\alpha$, the linear precoder achieves crosstalk-free performance. However, there is a degradation in performance if the channel deviates from being diagonally dominant. 
\end{tcolorbox}

\subsection{Legacy Techniques for Vectored VDSL System}
The extraordinary success of vectoring techniques for VDSL systems has proved pivotal in moving toward the next generation G.fast standard.  However, most of the techniques  developed for VDSL vectoring relied on the diagonal dominance of the channel to enable low enough implementation complexity \cite{Leung_Vectored_2013,Bergel_Signal_2013}. As discussed in Subsection III-B,  diagonal dominance does not hold over the G.fast frequency range. Nevertheless, in this subsection we briefly discuss these methods  to better depict the challenges of the transition from VDSL to G.fast.
\subsubsection{Crosstalk Cancellation  for VDSL}
The  works in \cite{cendrillon2006near, cendrillon2007} showed that the ZF based linear canceler/precoder yields performance close to the single user bound. 
In contrast to the wireless MIMO, it is possible to  bound the performance of the  ZF canceler  using the diagonal dominance  parameter. Thus, VDSL systems were derived using analytic estimates on the vectoring gain without the needing to rely on simulation/measurement results. The analysis for the ZF based cancelers  was further improved in \cite{zafar12spl} and \cite{bergel2012bounds}.  The bounds in  \cite{cendrillon2006near,cendrillon2007,zafar12spl,bergel2012bounds} prove that for strongly DD matrices, the noise enhancement in the upstream or power penalty in the downstream is negligible, and  performance is close to the optimal.

As shown in \cite{bergel2012bounds},  the data rate achievable by user $\ic$ in the upstream of a DSL system is  bounded:
\begin{align}{}\label{e: L1 upstream upper bound}
\begin{split}
\log_2 \left( 1+ \Gamma^{-1}P_{x,i} \sigma_{w,i}^{-2}  |\Hme_{\ic,\ic}|^2 \ffunc(\beta_{} ) \right)\leq	{\cal{R}}_i^{\rm zf} \\ \le  \log_2 \left( 1+ \Gamma^{-1}P_{x,i} |\Hme_{\ic,\ic}|^2\sigma_{w,i}^{-2}  (1+\beta_{\rm}) \right)
\end{split}
\end{align} where $\ffunc(\beta_{} )=\max\left\{0,1-2 \beta-\beta^2\right\}$. Thus, at lower frequencies where the DD property is significant ($\beta$ is significantly smaller than $1$ as depicted in Fig. \ref{fig:dd_measurement}) the ZF canceler is quite close to the upper bound, and hence near-optimal. On the other hand, these bounds are meaningless for $\beta>0.42$, and hence, we have no guarantee for the performance of the ZF canceler in G.fast. Similar performance was found for the downstream ZF system \cite{bergel2012bounds}.

Interestingly, as shown in Fig. \ref{fig:rate_equal_10users_212mhz} above, although the bounds become meaningless, ZF performs quite well  for G.fast as well. Thus, in many G.fast scenarios we can still use ZF, and use simulations/measurements to check its performance.

At this point, it is important to note that the complexity of ZF was too complex at the time, and the implementation of VDSL systems required another step using the DD property. Following \cite{leshem2007} the implementation of VDSL systems relied on a low complexity approximation of the ZF which was shown to be good when $\beta$ is small enough. Fig. \ref{fig:rate_equal_10users_212mhz}  shows that above $250$ \mbox{m}  performance using first oder approximation is equivalent to all other techniques. This approximation is described in the following subsection.
\subsubsection{Approximate ZF Canceler}
We can express  the channel matrix as $\mathbf{H}= \mathbf{D}(\mathbf{I}+\mathbf{D}^{-1}\mathbf{E})$ where $\mathbf{E}$ is the matrix containing the off-diagonal elements of $\mathbf{H}$.  The power series expansion of $(\mathbf{I}+\mathbf{D}^{-1}\mathbf{E})^{-1}$ is convergent  when the eigenvalues of the matrix $\mathbf{D}^{-1}\mathbf{E}$ are less than one.  Leshem  and Li \cite{leshem2007}  used the DD property of the channel to show  that  a first order approximation of the ZF $\mathbf{F}_{\rm azf}=(\mathbf{I}-\mathbf{D}^{-1}\mathbf{E})\mathbf{D}^{-1}$ achieves near-optimal capacity.  The complexity reduction is significant in that the matrix inverse operation is replaced
by the inverse of a diagonal matrix, which requires only $N$ single element inversions and $\mathbf{D}^{-1}$ is computed  anyhow using the FEQ (frequency domain equalizer) coefficients.
This is important since  ZF requires a channel matrix inversion at each tone. Such matrix inversions are frequently required due to changes in user status or variations in crosstalk characteristics, and hence cause an increase in overhead on the computational cost of the ZF receiver. 

Unfortunately, this approximation turned out to be useful only when $\beta$ is very small, and hence does not carry over well to G.fast systems (see Fig. \ref{fig:rate_equal_10users_212mhz}).
To avoid the channel matrix inversion, iterative algorithms for crosstalk cancellation in DSL systems have been proposed \cite{Dai2002,zafar2011}. However, these schemes do not perform well at higher frequencies for G.fast systems and introduce latency.  G.fast systems need to adopt more complicated algorithms whose design considerations  are further detailed in Section V.
\def\myDownArrow{\smash{
		\begin{tikzpicture}[baseline=-4mm]
		\useasboundingbox (-1,0);
		\node[single arrow,draw=red,fill=red!10,minimum height=4.6cm,shape border rotate=90] at (0.8,-3.5) {};
		\end{tikzpicture}
}}
\def\myUpArrow{\smash{
		\begin{tikzpicture}[baseline=1mm]
		\useasboundingbox (-2.2,0);
		\node[single arrow,draw=green,fill=green!10,minimum height=4.6cm,shape border rotate=90] at (-2.00,-1.40) {};
		\end{tikzpicture}
}}

\begin{table}[t]
	
	\centering
	\renewcommand{\arraystretch}{1.3}
	\caption{ A Summary on Crosstalk Cancellation Techniques}
	\label{table:crosstalk}
	\begin{tabular}{| p{2cm} || p{0.8cm} p{1.0cm}||  p{3.2cm}| }
		\hline
		Algorithm  & Performance& & Comments \\
		\hline
		ZF-THP \cite{ginis2002}  & $\myDownArrow$ & & Used in downstream. Negligible power penalty due to the modulo operation. Requires ordering to optimize performance.\\  \cline{1-1} \cline{4-4} 
		ZF-GDFE \cite{ginis2002} \cite{chen2007optimized}   & \multirow{8}{*}{\rotatebox[origin=c]{90}{\centering Data Rate}} \myUpArrow& \multirow{8}{*}{\rotatebox[origin=c]{90}{\centering Complexity}}&Used in upstream. Requires ordering to optimize performance. Problem of error propagation in the feedback loop.
		\\ \cline{1-1} \cline{4-4} 
		MMSE \cite{Wahibi2009icc}& &&    Performs marginally better than the ZF. Requires knowledge of noise covariance.
		\\ \cline{1-1} \cline{4-4} 
		ZF \cite{cendrillon2006near}\cite{cendrillon2007} &&  & Requires channel inversion.  Near-optimal for lower frequencies. 
		\\ \cline{1-1} \cline{4-4} 
		AZF\cite{leshem2007} &&    &  Does not require channel inversion. Near-optimal for longer loops and lower frequencies. Not suitable for typical G.fast applications. \\ 
		\hline
	\end{tabular}
	
\end{table}

\begin{tcolorbox}[colback=green!5!white,colframe=blue!75!black,fonttitle=\bfseries,title= D.   Crosstalk Cancellation using Approximate Inversion ,breakable]	
	To illustrate the approximate inversion method, we extend the model in Box A  to a three user model: 
	\[ \left[ \begin{array}{c}
	Y_1 \\
	Y_2\\
	Y_3
	\end{array} \right]
	= 
	H_d\left[ \begin{array}{ccc}
	1 & \alpha&\alpha \\
	\alpha & 1&\alpha\\
	\alpha & \alpha&1
	\end{array} \right]
	\left[ \begin{array}{c}
	X_1 \\
	X_2 \\
	X_3
	\end{array} \right]
	+\left[ \begin{array}{c}
	W_1\\
	W_2\\
	W_3
	\end{array} \right].
	\tag{D1}
	\]
	\hspace{+0.3cm} 	We denote $\mathbf{D}$ as a diagonal matrix of $\mathbf{H}$, and $\mathbf{E}$ as the matrix containing the off-diagonal elements of $\mathbf{H}$. The first order approximation method approximates the inversion of the ZF canceler:
	\[ 
	\mathbf{F}_{\rm azf} =(\mathbf{I}-\mathbf{D}^{-1}\mathbf{E})\mathbf{D}^{-1}=
	\frac{1}{H_d}\left[ \begin{array}{ccc}
	1 & -\alpha & -\alpha\\
	-\alpha &1 &  -\alpha\\
	-\alpha &-\alpha &  1
	\end{array} \right].
	\tag{D2}
	\]
	This canceler does not remove the crosstalk completely since $\mathbf{F}_{\rm azf}\mathbf{H}$ is not identity.  Applying the canceler  $\mathbf{F}_{\rm azf}$ on the received signal results in the signal estimate:
	\begin{align}
	\begin{split}
	\hat{X}_1 = (1-2\alpha^2)X_1-\alpha^2X_2-\alpha^2X_3+\frac{1}{H_d}W_1-\\
	\frac{\alpha}{H_d}W_2-\frac{\alpha}{H_d}W_3,
	\end{split}
	\tag{D3}
	\end{align}
	and the spectral efficiency:
	\begin{align}
	{\cal{R}}^{\rm azf} = \log_2\left(1+\rm{SNR}_{\rm awgn}\frac{(1-2\alpha^2)^2}{2\alpha^4\rm{SNR}_{\rm awgn}+2\alpha^2+1}\right).
	\tag{D4}
	\end{align}
	\hspace{+0.3cm} 	Thus, the approximate ZF is close to optimal if both $\alpha\ll 1$ and $\alpha^4 {\rm SNR_{awgn}} \ll 1$. This is much better than the no cancellation case, which $\alpha^2  {\rm SNR_{awgn}}\ll 1$. 	However, if $\alpha$ is not very small, and the SNR is high (as in most frequencies in G.fast)  the performance of the approximate ZF 	can be significantly lower than the exact ZF.
	
	\centering
	{\includegraphics[width=\linewidth]{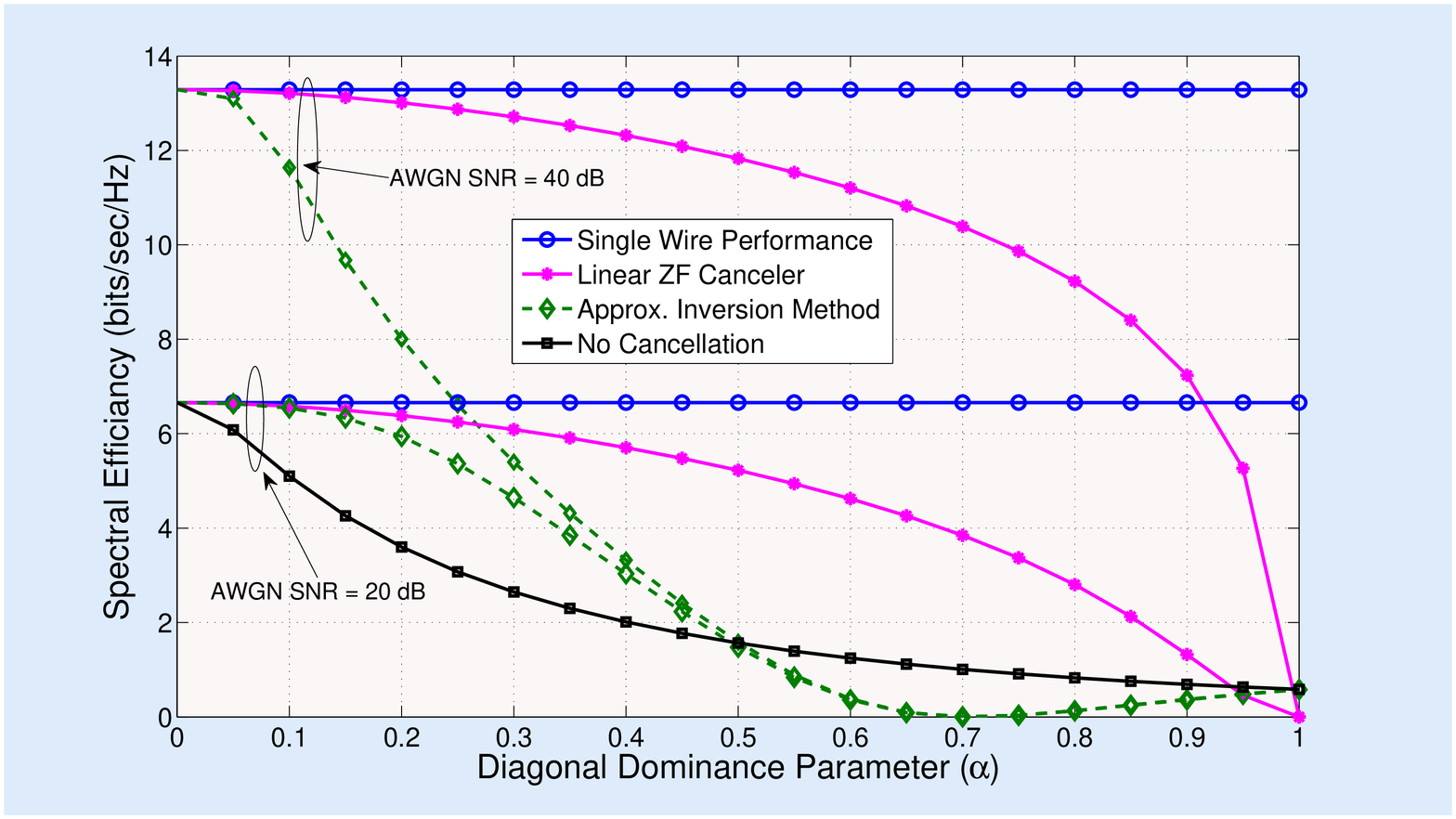}}
	\captionof*{figure}{Fig. S3: Performance of linear approximate inversion canceler versus DD parameter for three users with  AWGN SNR of $40$ \mbox{dB} and $20$ \mbox{dB}. For  the DD parameter $\alpha<0.1$ (which is typical of VDSL), the linear ZF canceler performs close to  AWGN performance.}
	
\end{tcolorbox}

\subsubsection{ Adaptive Crosstalk Cancellation}
Adaptation in the crosstalk cancellation schemes requires an update of the canceler coefficients based on the dynamics of the channel. In the upstream VDSL, there are several adaptive algorithms that minimize the mean square error (MSE) at the output of the crosstalk canceler. Certainly, the most popular algorithm is the LMS algorithm \cite{Widrow_Adaptive_1960}. LMS is a stochastic gradient descent method that under certain conditions converges to the MMSE solution. The popularity of the LMS algorithm is mostly due to its simplicity. The LMS crosstalk canceler (in each tone) can be described as a matrix
\begin{equation}
\label{eq:LMS_matrix}
{\bf F}[t]\triangleq\left[{\bf f}_{1}[t],{\bf f}_{2}[t],\cdots,{\bf f}_{_{N}}[t]\right]
\end{equation}
where ${\bf f}_{N}[t]$ denotes the canceler vector for line $N$ at time t.  Hence, the output of the LMS crosstalk canceler can be written as
${\bf z}[t]={\bf F}^H{\bf y}[t]$. For an error vector ${\bf e}[t] = {\bf x}[t]-{\bf z}[t]$, the LMS recursion can be written in a matrix form as
\begin{equation}\label{eq:LMS_matrix_recursion}
{\bf F}[t+1]={\bf F}[t]-2\mu{\bf y}[t]{{\bf e}^H[t]},
\end{equation}
where $\Fm(t)$, $\yv(t)$ and $\mathbf{x}(t)$ denote the value at time $t$ of the precoding matrix, the received symbol and modulated data, respectively. With an appropriate selection of step size $\mu$, the LMS is guaranteed to converge. Nevertheless, its convergence rate depends on the eigenvalues of the input correlation matrix. Thus,  LMS based adaptive algorithms are very efficient and converge quite rapidly for diagonally dominant VDSL channels. On the other hand, these algorithms  become inefficient or converge very slowly at higher frequencies due to bad conditioning of the input correlation matrix. Hence, the conventional LMS algorithm is not suitable for the G.fast system and new methods are needed.

For the VDSL system, the adaptation of a crosstalk canceler matrix in the downstream is more complicated because  the transmitter at the optical network unit (ONU) cannot directly measure the channel matrix. Hence, the calculation of the precoding matrix must rely on feedback from the receivers. Two main types of feedback have been considered: channel estimation feedback and signal error feedback. Channel estimation feedback is based on the transmission of orthogonal (synchronized) pilot symbols from all transmitters simultaneously, and an estimation of a row of the channel matrix by each receiver. This estimated row is then transmitted (through the upstream) back to the transmitter, which uses it to construct the full channel matrix $\mathbf H$ and to calculate the precoding matrix. 

Signal error feedback is a more DSL-specific method, and is based on the feedback of a quantized version of the error signal measured by each receiver\cite{louveaux2006}. The ONU groups all the error signal feedback from each CPE into a vector to adapt the precoder coefficient matrix. A simpler version of an adaptive precoder was proposed independently by Louveaux and van der Veen \cite{louveaux2010} and by Bergel and Leshem \cite{bergel2010}. The adaptive precoder  was 
shown to be  very robust, and  converged in all channels with a strong RWDD property  \cite{Binyamini_Arbitrary_2012} \cite{idobergel2013}. However, these approaches lost their attractiveness when the new G.fast standard adopted a TDD approach. Thus, in G.fast, the ONU can directly measure the channel in the upstream phase and use its estimation in the downstream phase with no need for feedback.

\section{Design Considerations and Challenges for  G.fast }
The G.fast transmission technology is under development and has different features than the current vectored VDSL technology.  The techniques  developed for the VDSL system thus cannot simply be applied to the G.fast system. Yet, efficient design methodologies for the G.fast system are required to deliver data at the rate of $1$ Gbps per user over short telephone lines. Recently, various novel techniques have been developed for the G.fast system. These techniques are still not sufficient to reach the desired rates with currently available hardware. In this section, we discuss various differences between G.fast and VDSL,  give an overview on recent research, and highlight the design considerations  for the G.fast system. We also point out some of the topics that are important for G.fast implementation, which have not been studied sufficiently and require further research. 

\subsection{G.fast Characteristics}
We first present a concise representation of the main G.fast characteristics, as a complement to the general system model given in  Section II.   G.fast transmission model consists of $N$  vectored users (typically upto $100$) and operates over loop lengths shorter than $250$ \mbox{m}.  The G.fast standard targets an aggregate data rate (combined upstream and downstream) of $1$ \mbox{Gbps} per user. The data rate performance of  G.fast system is limited by the FEXT since  NEXT is eliminated with the TDD scheme which provides  independent transmissions on upstream and downstream directions over whole bandwidth. The transmission bandwidth starts at $2.2$ \mbox{MHz} and ends at $106$ \mbox{MHz} for the low bandwidth version of G.fast and $212$ \mbox{MHz} for the  upcoming version.  The system employs DMT modulation of size $K$ sub-carriers of width $51.75$ \mbox{KHz} for each user corresponding to a total $K=2048$ sub-carriers for $106$ \mbox{MHz} G.fast  and $K=4096$ sub-carriers for  $212$ \mbox{MHz} G.fast system.   The  channel matrix is  diagonally-dominant at the lower frequencies, but not at the higher frequency tones. Each user transmits a QAM symbol (with a rate of $48000$ symbols per second) of unit energy with a gain scaling factor to control transmission power. The per-tone PSD mask is based on the frequency of operation: $-65$ \mbox{dBm/Hz} for  $f\leq 30$ \mbox{MHz}, $-76$ \mbox{dBm/Hz} for  $30$ \mbox{MHz}  $< f\leq 106$ \mbox{MHz}, and  $-79$  \mbox{dBm/Hz} for  $f> 106$ \mbox{MHz}. Each transceiver has a total  maximum transmit power of $4$ \mbox{dBm}. The additive noise is AWGN with a PSD of  $-140$ \mbox{dBm/Hz}.   The target bit-error-rate (BER) is set to $10^{-7}$ and an SNR gap is $\Gamma=9.75$  \mbox{dB}. The specified  noise margin  of $6$ \mbox{dB}  and coding gain of  $5$  \mbox{dB} lead to a transmission gap of $10.75$ \mbox{dB}. The G.fast limits bit loading to $12$ bits per tone i.e.  QAM constellation size of up to $4096$ points.

\subsection{Channel Coherence Time}
The DSL channel  is a slowly time-varying  since it consists of twisted-pair copper wires in a static cable binder with fixed user terminals. The time variability can be attributed mainly to   changes in customer wiring and  temperature variations on the time scale of a few minutes or more.   Therefore, the DSL channel has a long coherence time  (defined as the time in which there is no effective variation in the channel impulse response). This has several advantages as regards DSL system design in contrast to wireless communication systems. Furthermore, it allows multi-user operation with as many as $100$ users simultaneously, an order of magnitude larger than existing MIMO wireless systems.  The long coherence time allows for  almost perfect CSI acquisition. This enables the robust design of crosstalk cancellation schemes. In addition, tracking and updating of the channel matrix is less frequent, which reduces the overhead required for pilot symbols inserted between the data symbols.  With large channel coherence, iterative algorithms for power allocation as well as crosstalk cancellation are feasible even with a very large number of lines.

\subsection{Channel Estimation and  Calibration}
As discussed above, the long coherence time enables the receivers to obtain very good channel estimates, which are important for cross talk cancellation. Hence, channel matrix estimation in upstream transmission is much easier than in the downstream. In upstream transmission, users transmit known training  symbols and  a simple least squares technique can be used to estimate the channel matrix at the distribution point. The length of each user's training symbols should be at least equal to the  number of vectored users. In the downstream, the channel estimation is carried out at the user terminals where each row of the channel matrix is  estimated. This increases the complexity of the users' modem  and power consumption. However, the main challenge is to forward the  estimated channel from each user to the distribution point.

In VDSL systems, which are based on the FDD scheme, the channel coefficients are quantized at each user and transmitted to the DP over the upstream channel. This increases the overhead and  complicates the design with a reduction of  quality of CSI for transmit precoding.  The TDD duplexing scheme of G.fast makes it possible   to exploit channel reciprocity  to avoid the complicated feedback protocol.  This also enables the DP to perform all CSI related tasks. Using channel reciprocity,  the downstream channel  can be estimated as $\tilde{\mathbf{H}}_{d} = \tilde{\mathbf{H}}^{T}_{u}$ where  $\tilde{\mathbf{H}}_{u}$ is the estimated upstream channel.  In practice, the transmit and receive paths at the DP are not identical;  hence, the estimated downstream  channel needs to be calibrated to compensate for this mismatch. This calibration process is still in development and has not yet received sufficient academic attention.

\begin{figure}[t]	
	\begin{center}
		\includegraphics[width=\columnwidth]{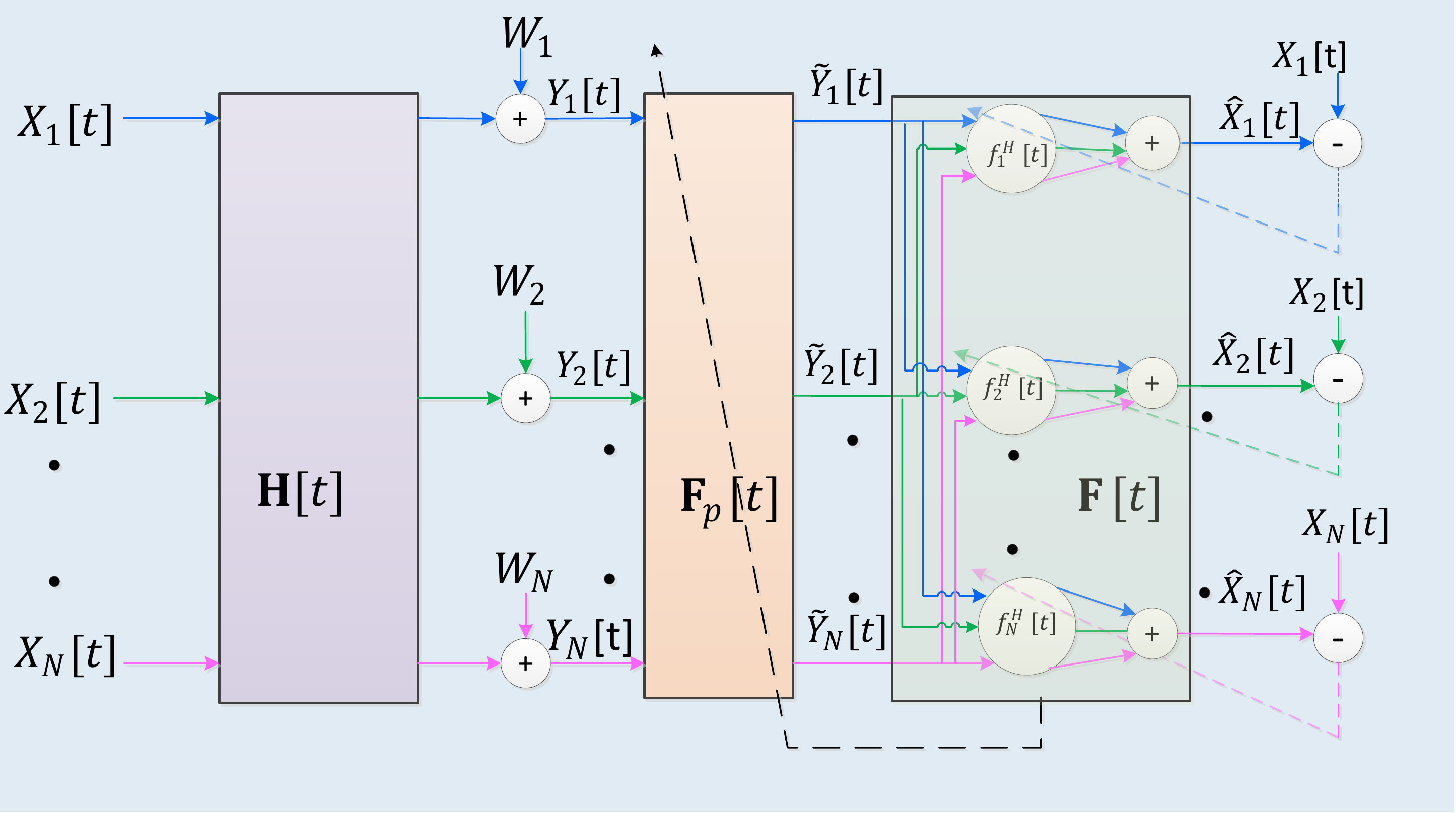}
	\end{center}
	\caption{A two-stage LMS algorithm for adaptive crosstalk cancellation in upstream G.fast.}  
	\label{fig:deeplms}
\end{figure}

\subsection{Adaptive Crosstalk Cancellation}
The adaptation of a crosstalk canceler matrix to track the channel dynamics is another important design consideration for the G.fast system.   In general  traditional adaptive algorithms designed  for the VDSL  system can be applied to the G.fast system. However, these algorithms and mainly the LMS algorithm \cite{duvaut08} become inefficient or converge very slowly at higher frequencies due to bad conditioning of the input correlation matrix.  At higher frequencies the received signal's covariance matrix is badly conditioned due to loss of diagonal dominance characteristics. There is an ongoing research on adaptive techniques for upstream transmissions.

As an interesting new paper, \cite{zanko2016eusipco} \cite{Avi_deeplms_arxiv2017}  presents a novel algorithm that can speed up the convergence of the LMS crosstalk canceler by preprocessing  the input received signal with a judicially designed matrix preprocessing matrix, as shown in Fig.~\ref{fig:deeplms}. The proposed  two-stage LMS algorithm first preprocesses the input signal by multiplying it with a matrix $\mathbf{F}_p[t]$ and then applies a standard LMS on the preprocessed signal that updates another decoding matrix, $\mathbf{F}[t]$.  The conventional LMS is accelerated by updating (at carefully selected times) the preprocessing matrix to include also the LMS matrix:   ${\bf F}_{{p}}[t+1]= {\bf F}_{{p}}[t]{\bf F}_{\rm{}}[t+1]$ and LMS $\mathbf{F}[t]$=
$\mathbf{I}$.  The paper shows that the updates of the preprocessing matrix   speed-up the
convergence of the LMS crosstalk canceler by reducing the condition number of the correlation matrix at the LMS input.  Since, the preprocessing matrix is not frequently updated, the complexity of the algorithm is approximately twice the complexity of the conventional LMS.

For the downstream, the TDD scheme in the G.fast facilitates a simpler approach to adaptive algorithms than the VDSL. Using  channel reciprocity,  feedback from the users is no longer necessary. Instead, the adaptive algorithm can be carried out solely in the upstream, and the resulting FEXT cancellation matrix is guaranteed to also be a good precoding matrix for the downlink.

\subsection{Optimized Crosstalk Cancellation Schemes}
The performance of crosstalk cancellation techniques (as described in Section IV) depends mainly on  the   characteristics of the DSL channel.  Since the G.fast channel is not diagonally dominant at higher frequencies, the crosstalk cancellation schemes  for the VDSL system are no longer optimal for most of the G.fast channel bandwidth. We overview recent research works that describe novel approaches to deal with strong crosstalk in the G.fast system \cite{Muller2014icc, ZUTHP2016, hekrdla2015,strobel2015icc,strobel2015gc,Barthelme2016gc}.

\subsubsection{Power Normalization}
Use of  linear cancellation in the downstream transmission requires an energy normalization that can guarantee the power constraint for all lines. For the VDSL (low) frequencies, it was shown that simple power normalization results in near optimal performance \cite{cendrillon2007}. But, this is not the case for G.fast frequencies, and power normalization can significantly reduce the SNR   when the channel matrix is ill-conditioned.  A single normalization factor $\beta_{\max} = \max_i \sum_{j=1}^N|{F}^{}_{ij}|^2$ (i.e. gain scaling  $\sqrt{\mathbf{G}_{}/\beta_{\max}}$) has been  shown near optimal for low frequencies \cite{cendrillon2007}.

A simple power normalization approach, \cite{maes2014icc} adopted a two stage normalization. In the first stage, the power of each user is normalized according to its overall line power, while the second stage is the simple overall power normalization of VDSL. A more effective approach, \cite{Muller2014icc} optimizes the per-line power normalization scheme via a linear program  that further improve the performance and balances the rates among the users.

\subsubsection{Ordered Successive Interference Cancellation }
The non-linear interference cancellation schemes for the  downstream  does not  require power normalization as their power increase is negligible (e.g., because of the modulo operation in THP). However, the issue of  user ordering for successive interference pre-coding becomes very important.  As the channel gains  depend on the decoding order, the users precoded earlier experience higher data rates, while the users precoded later achieves  lower data rates creating undesirable rate variance among users. 
Several works (e.g., \cite{ZUTHP2016}, \cite{hekrdla2015}) have suggested novel approaches for user ordering that both improve the total throughput and reduce the rate variance.  In the upstream, the ordering affects the  performance of the ZF-GDFE, which strongly  depends on the detection ordering due to the error propagation  effect \cite{chen2007optimized}. 

Comparing to wireless communications, the G.fast requires more complicated processing as it simultaneously serves more user and with higher user rates. On the other hand, the low rate of change in the channel and in the user activity allows a longer processing time for this optimization problem. The problem becomes even more complicated as we take the whole bandwidth into account, and try to jointly optimize the complete network. This is discussed in the following subsection.

\subsection{Joint Multi-User Crosstalk Cancellation and DSM }
Dynamic spectrum management (DSM) techniques were shown to improve the performance of DSL systems (e.g., \cite{Yu2002DSM, Cendrillon2006TCOM, Tsiaflakis2008tsp, Huberman2012_DSM}) by improving the resource allocation between all users over the system bandwidth.   DSM have evolved mostly independently  of multiuser crosstalk cancellation and both were considered (separately) as different level of coordination in  DSL systems \cite{ginis2002} \cite{cendrillon2006near}.

Novel research approaches for G.fast consider a joint signal and spectrum coordination. The joint implementation of crosstalk cancellation and spectrum management can be formulated as an optimization problem that need to    derive  the optimal precoder and decoder and optimal power allocation subject to a given set of  constraints. Practical G.fast systems typically face three power constraints: The first constraint is the PSD mask $P_{k, \rm mask}^{}$ which represents the maximum PSD allowed at tone $k$. This constraint is defined by regulation and prevents G.fast from allocating too much power in certain frequency band.  The second constraint is transmit power per-line $P_{\rm line}$ to account for the limited range of transmit amplifiers. The third constraint is the bit cap $b_{\rm cap}$ for bit-loading and reflects the limited capabilities of the modulator, which does not allow constellation sizes greater than a threshold.  

It is noted that similar optimization problems have been studied extensively for wireless networks \cite{Boccardi2006,Yu2007,Shi2008,Christensen2008}. However, proposed algorithms from wireless systems are not directly applicable to  G.fast context due to the use of large number of users in the DSL systems as well as additional constraint of per-line power $P_{\rm line}$ in addition to the  PSD mask $P_{k, \rm mask}^{}$ (i.e. existing per-antenna power constraint). 
The joint sum-rate optimization of the precoder and the powers can be obtained for example as \cite{strobel2015gc}:\begin{maxi}
	{\mathbf{F}_{k},\mathbf{p}_{k}}{\sum_{n} \sum_{k} \mathbf{\psi}_{k,n}(\mathbf{F}_k, \mathbf{p}_{k}), \forall k}{}{}
	\addConstraint{{\rm diag}(\mathbf{F}_{k}\mathbf{F}_{k}^{H})}{\leq \mathbf{p}_{k,\rm mask}^{}}{, \forall k}
	\addConstraint{\sum_{k}{\rm diag}(\mathbf{F}_{k}\mathbf{F}_k^{H})}{\leq \mathbf{p}_{\rm line}^{}}{, \forall k}
	\addConstraint{\mathbf{\psi}_{k,n}(\mathbf{F}_k, \mathbf{p}_{k})}{\leq {b}_{\rm cap}^{}}{, k=1,\cdots K,  n=1,\cdots N}
	\label{eq:optimization_one}
\end{maxi}
where $\mathbf{\psi}_{k,n}(\mathbf{F}_k, \mathbf{p}_{k})$ is the achievable rate for the line $n$ and tone $k$ for a given set of power allocation vectors $\mathbf{p}_{k} \in \mathbb{R}^{1 \times N}$ and  precoding matrix $\mathbf{F}_k  \in \mathbb{C}^{N \times N}$ in the desired interference cancellation scheme. The vectors  $\mathbf{p}_{k, \rm mask} \in  \mathbb{R}^{1 \times N}$  and  $\mathbf{p}_{\rm line} \mathbb{R}^{1 \times N}$ denotes the PSD masks and  per-line powers for $N$ users at tone $k$, respectively.

The case of MMSE precoding was solved using a closed form solution for the precoders given the power allocation per tone per line, and using the uplink-downlink duality \cite{strobel2015gc}. Recently the MMSE precoding has been proposed in the non-linear THP framework \cite{Barthelme2016gc}. The algorithms to solve these optimization problems are quite complex for G.fast systems.  However, these techniques are useful as they provide a computable bound to compare computationally efficient algorithms for other optimization problems.

To reduce the complexity of algorithms, the DSM algorithms can be applied with a fixed canceler structure such that the transmit power optimization minimizes the residual crosstalk and improves the performance. It was shown  \cite{ginis2002},  \cite{cendrillon2006near} that the application   of ZF-GDFE or linear ZF canceler  decouples  the power  allocation problem  and essentially becomes independent for each user. In this scenario, an iterative water filling (IWF) algorithm can be applied for each user to maximize their own data rate selfishly. This results in sub-optimal performance but gives  a low complexity  solution that can be readily implemented. Current research aims to find better algorithms to fill the gap between the infeasible optimal solution and the simplified ZF-IWF solution.

\subsection{Design consideration for Crosstalk Cancellation Schemes}

\begin{figure}[t]	
	\begin{center}
		\includegraphics[width=\columnwidth]{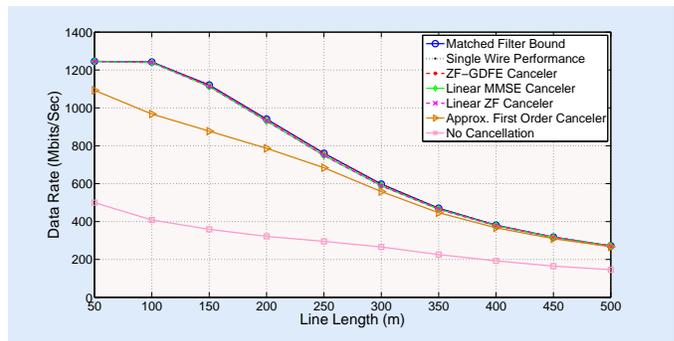}
	\end{center}
	\caption{Average achievable user rate over the whole bandwidth of $106$ MHz vs. binder length. The binder is composed of $10$ users with equal line lengths.}
	\label{fig:rate_equal_10users_106mhz}
\end{figure}
\begin{figure}[t]	
	\begin{center}
		\includegraphics[width=\columnwidth]{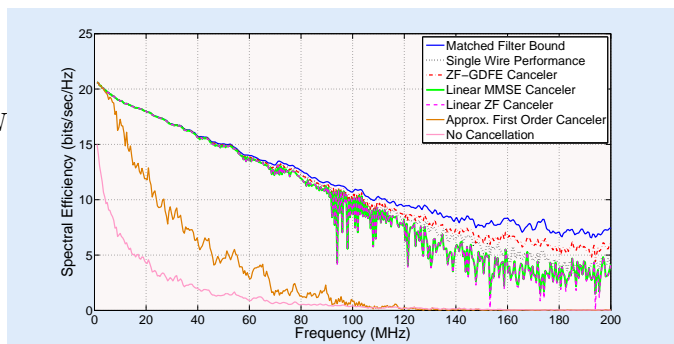}
	\end{center}
	\caption{Spectral efficiency performance using channel measurement \cite{BT} for a  binder with $10$ lines of $100$\mbox{m} length. }
	\label{fig:spectral_efficiency_measu}
\end{figure}
\begin{figure}[t]	
	\begin{center}
		\includegraphics[width=\columnwidth]{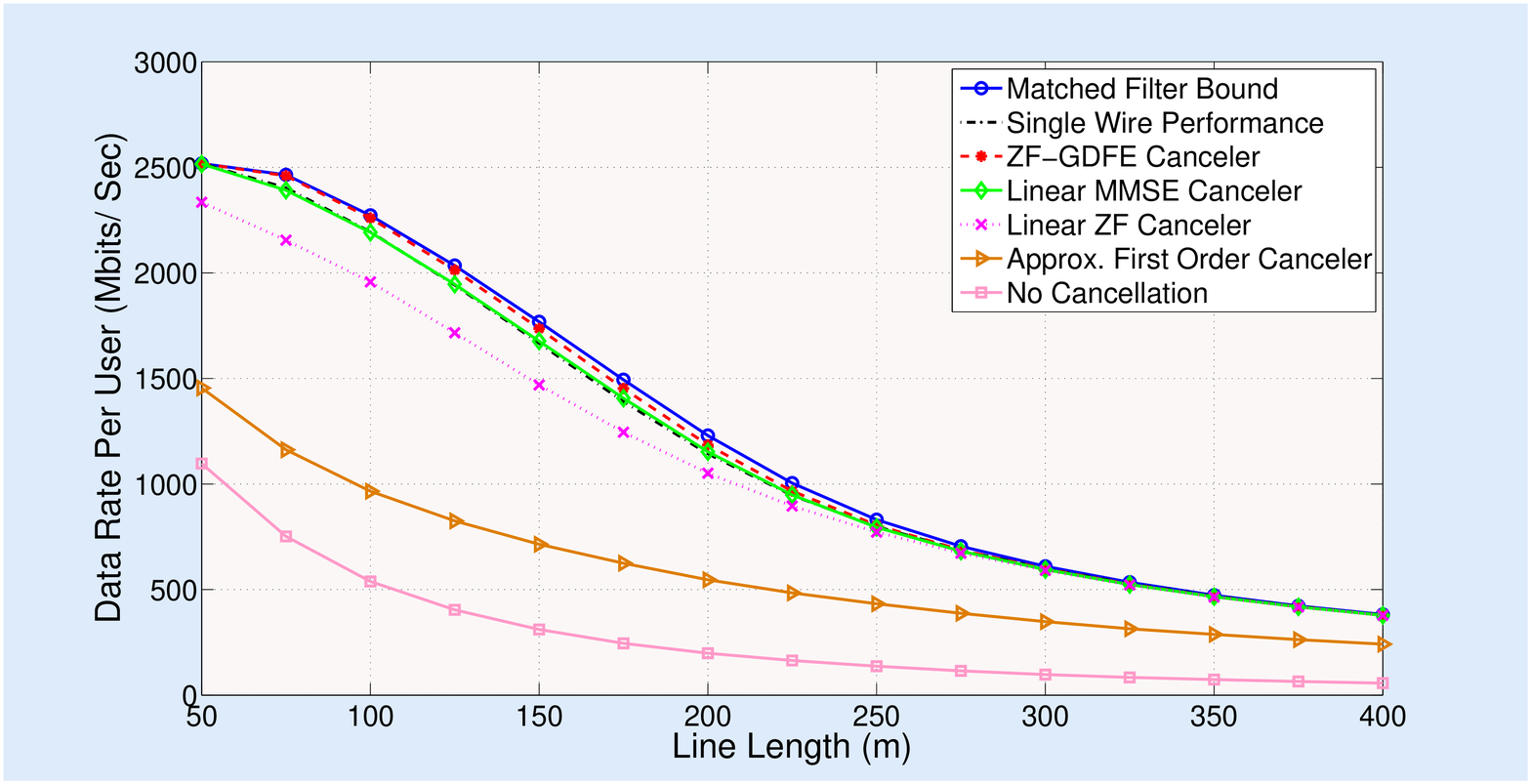}
	\end{center}
	\caption{Achievable data rates of $15$ users in a  binder with uniformly spaced line lengths. All users use G. fast $212$ MHz with various cancelers.}
	\label{fig:simulation_gfast_rate_cancelers}
\end{figure}
The most important design issue in G.fast is the crosstalk cancellation scheme. As  discussed in Section \ref{section:crosstalk_techniques} there is a performance-complexity tradeoff, where linear cancelers are simpler, but lose on performance. We now discuss the choice of crosstalk cancelers as regards the performance over standard channels. Performance is evaluated using measured channel \cite{BT} as well as stochastic MIMO channel models with CAT5 cables \cite{g9701}. The measured channel contains data from a $0.5$ mm cable with $10$ pairs each measuring $100$ \mbox{m}.  For the rate-reach plot, we consider  a simulated channel with two scenarios:  the first scenario consists of $10$ lines each of equal length that range from  $50$\mbox{m} to $500$\mbox{m}, and the second scenario $15$ considers lines with different line lengths uniformly spaced  from $50$ \mbox{m} to  $400$ (intervals of $25$ \mbox{m}). The lowest frequency was $2.2$ \mbox{MHz} for both $106$ \mbox{MHz} and $212$ \mbox{MHz} G.fast systems which is defined in the standard \cite{g9701} for backwards compatibility with legacy ADSL systems. The noise PSD was $-140$ \mbox{dBm/Hz} and the transmit PSD specification at each tone was taken from \cite{g9700psd}. A bit cap of $12$ bits per tone (i.e., $4096$ QAM densest constellation) was maintained and an effective SNR to Shannon gap  of $10.75$ \mbox{dB} was used. 

The ITU-T  standard \cite{g9701} recommends linear crosstalk cancelers for the G.fast $106$ \mbox{MHz} since the loss of the linear scheme is marginal up to 106 MHz (see Fig. \ref{fig:rate_equal_10users_106mhz}). The rate-reach  in Fig.~\ref{fig:rate_equal_10users_106mhz} shows that the linear crosstalk schemes are near optimal for $106$ \mbox{MHz}.  Note that  at  very high frequency tones (above $90$ MHz), the diagonal dominance is not very strong, which degrades the performance of linear schemes as shown in Fig. \ref{fig:spectral_efficiency_measu}. However, at such  high frequencies, the spectral efficiency is significantly lower due to the signal attenuation and therefore,  the overall loss in data rate (computed over all tones) is marginal. The performance of  the linear approximate inversion is significantly degraded and is not recommended for G.fast.  It can be seen that the linear MMSE canceler performs marginally better than the ZF at higher frequencies; however, it is recommended since it provides extra protection against external noise and its complexity is almost identical to that of the ZF.
Recently,  an asymptotic analysis on the performance of ZF receiver using theory of large dimensional random matrices has been presented for G.fast system in \cite{zafaruddin_eilat2016}.

The  G.fast channel up to 212 MHz is not diagonally dominant and the linear ZF scheme is no longer near-optimal for  crosstalk cancellation (see Fig. \ref{fig:simulation_gfast_rate_cancelers} ). Here, non-linear cancellation schemes are preferred,  but the MMSE linear scheme still gives reasonable performance, with at most a $5$\% degradation compared to the MFB and $3$\% compared to ZF-GDFE. The non-linear schemes are computationally more complex than the linear techniques. However, their  performance can be improved with proper ordering procedures among users \cite{hekrdla2015}. Recent works \cite{maes2016} show that the impact of the channel estimation error on the non-linear is more crucial on the G.fast system than the VDSL system. This is also an important design consideration for the non-linear G.fast system.

It is noted that the above simulation results have been obtained without optimal power control (as discussed in the  subsection V-F). Algorithms for joint DSM with crosstalk cancellation in \cite{strobel2015icc}, \cite{strobel2015gc} showed that there was a significant performance improvement for the optimized linear ZF and non-linear ZF-GDFE at higher frequencies. However,  the considered algorithms have higher  computational complexity due to  the large-dimensional power allocation  in G.fast systems.

\subsection{Implementation Complexity}
Crosstalk cancellation schemes require significant computational resources given  various signal processing operations at a large scale. The first  is the memory requirement of the processor to store $N^2$ channel coefficients per tone. Since the number of tones is very large, $K=4096$, the memory requirement is significant. The efficient usage of memory for this task also needs further investigation. For example, the memory requirement can be reduced if the channel coefficients are  stored at a few tones throughout the band and interpolated on the other tones. However, this will reduce the data rate performance because the error in channel estimates increases with the interpolation interval. These tradeoffs are standard  signal processing design. The loss due to quantization and interpolation can be characterized using the techniques of \cite{sayag2009}.

Another complexity consideration is the computation of the canceler/precoder matrix. The linear ZF and MMSE canceler/precoder require inversion of many channel matrices whose complexity is very large $\mathcal{O}(KN^3)$. The QR decomposition in the ZF-GDFE canceler/ THP precoder  requires similar high computational resources. The approximate inversion method \cite{leshem2007} which avoids the matrix inversion is not recommended for the G.fast system at higher frequency tones.

The adaptation of the canceler/precoder coefficients should be simple and  not need costly computation. The G.fast standard provisions a single sync symbol after every $274$  data symbols to track the channel. The authors in \cite{duvaut08} have proposed a  scheme to directly  track the canceler matrix whose convergence depends on the diagonal dominance parameter of the channel. Further research is required to derive computationally efficient adaptive canceler schemes for the G.fast channel. 

However, the major complexity occurs when the $N \times N $ canceler/precoder matrix is applied on the received/transmit vector of size $N$. This requires $KN^2f_s$ instructions per second which becomes prohibitively high at $1.9$ GIPS (a billion instructions per second) for a $N=10$ users binder  operating at $f_s=48$ \mbox{KHz} for $4096$ tones.

\subsection{G.Fast Deployment and Coexistence}
With the penetration of fibers closer to the end users, various deployment scenarios  have been considered for the G.fast fiber to the distribution point (FTTdp) system.  The FTTdp consists  scenarios where the DP is either  mounted on a pole or placed  underground at the curb, building, basement, etc. The G.fast FTTdp fiber node can be as far as $250$ \mbox{m} to serve dozen of users and as close as $50$ \mbox{m} to serve a single user. Since such a scenario requires the deployment of many new DPs, the G.fast also considers the possibility that the DP will be powered from the customer's CPE. This technology known as reverse powering eliminates the need for a power infrastructure for the DP and hence lowers the deployment cost. This technology also introduces  new challenges  such as use of a single wire for both broadband and power transmissions and issues related to safety guidelines for  power transmission in a home network.

Another issue affecting the G.fast deployment is the overlap of its bandwidth with other radio  and broadcast services. Radio services  overlap a few consecutive tones  with the G.fast system throughout the band. These tones need to be either masked or notched such that no interference is experienced by the radio service. Broadcast radios such as FM  are broadband services and  cover a large bandwidth ($20$ \mbox{MHz}). The G.fast system has to avoid data transmission over these bands, which limits the rate performance of G.fast. Note that these unused tones can be employed for system provisioning. It should be emphasized that efficient transmissions methods are used to enable the co-existence of  the G.fast system  with  other broadband technologies without disturbing the band profile.

The coexistence of the G.fast system with legacy lines is another important issue  \cite{medeiros2014icc}.  Vectored VDSL systems have already been deployed by various service providers  and may coexist with  G.fast users in the same binder. VDSL fall back and coexistence is necessary for mass deployment of G.fast. Most service providers require that migration from VDSL to G.fast should be seamless, such that the G.fast must be able to peacefully co-exist with VDSL2 from the same DP. A common approach to co-existence is to have both technologies  use non-overlapping spectra. This is a simple approach to avoid any interference between two technologies. However, leaving the lower frequency bands where the spectral efficiency is very high limits the data rate performance of the G.fast technology. The coexistence of G.fast and VDSL in the same DP with an overlapping spectrum is a challenging problem, somewhat similar to the problems faced by  ADSL deployment in Japan parallel to TDD based HDSL \cite{g9923}.

\subsection{Crosstalk Cancellation with Discontinuous Operation}
DSL access networks boast ``always on" as a distinguishing feature of their broadband service. However, a typical DSLAM (at the DP) is fully on but does not transmit data all the time,  yet most DSL lines do not switch off. Exploiting these low-power modes can be highly effective in creating  a “green”  access network.   Moreover, energy efficiency is very important for the G.fast DP because they do not have a local power supply, but rather are reverse powered  from the subscribers via the copper wires. Discontinuous operation basically means that no data symbols are transmitted when there are no data available. While a VDSL system sends idle data packets, the G.fast system can mute these data symbols and switch off the analog front-end components to enhance energy efficiency \cite{maes2014icc}.

Crosstalk cancellation in conjunction with discontinuous operation has become challenging  \cite{strobel2016}. A concise problem formulation on the precoded discontinuous process has been presented in \cite{maes2014icc}. Here crosstalk cancellation must be maintained on active lines, while other lines are  in the discontinuous mode.  Novel algorithms are required to adapt the linear and non-linear precoder sub-matrices corresponding to active lines  in order to enjoy   energy savings as well as crosstalk cancellation.

\section{Discussion and Future Directions}
In this article, we provided an overview of multiuser signal processing techniques for crosstalk cancellation in the next generation G.fast system. We discussed the salient features of this upcoming DSL technology and highlighted the key differences from its predecessors. This tutorial article shows that considerable advances have been  made in recent years for crosstalk cancellation in vectored DSL systems. However, the G.fast system poses new challenging problems and   further research is still needed to fully realize the goal of  achieving gigabit date rates over telephone channels in access networks.

The use of wider bandwidth and the TDD duplexing scheme, among others, are the two most distinguishing features in the G.fast  that require special attention from the signal processing community. The DSL channel has many specific characteristics that differ from  most of the wireless channels. These unique characteristics allowed the implementation of multiuser interference cancellation possible for large number of users. The G.fast channel still integrates  these characteristics but loses a key one at higher frequencies: it is no longer diagonal dominant.

We showed that the approximate ZF precoder/canceler, which was one of the enabler of massive vectoring, does not perform well enough at the G.fast frequencies.  We provided an overview of the most important techniques (linear and non-linear) and explained them through simple examples.  However, all these techniques require high implementation complexity, and more research is required to make multiuser processing work on DSL channel up to $212$ MHz.

Similarly, we showed that the adaptive schemes designed for the VDSL system do not perform well  on the G.fast channel. Additional research is required to address the convergence of  these algorithms,  in addition to efficient calibration processes to use  channel reciprocity in the downstream transmission.

Additional issues that require more research include the study of mitigation techniques for interference from uncoordinated lines \cite{zafar_iet2015} and noise external to the binder on the vectored 
G.fast system \cite{zafar_icc2016}, dynamic spectrum management, and better channel modeling.  So far the channel modeling at the higher frequency is less accurate than what we have for VDSL. More measurement campaigns and additional statistical studies are required to improve this modeling. Furthermore, the characterization of the crosstalk channel is just in its early steps and very few publications have addressed it \cite{van2011}, \cite{van2012} \cite{Brink2017}. As G.fast will heavily rely on crosstalk cancellation, a good stochastic modeling is required for developing and testing of G.fast algorithms.

The use of G.fast system as an alternative backhauling option for the $5$\mbox{G} networks could keep the copper in gold category.

\bibliographystyle{IEEEtran}
\bibliography{spm_bibtex_file}

\small
\section*{Authors' Biographies}
\begin{itemize}
	\item \textbf{S.~M.~Zafaruddin} (smzafar@biu.ac.il)
	received the B.Tech. degree from Jamia Millia Islamia University, New Delhi, India, in 2003 and M.Tech. degree  from Kurukshetra University, Kurukshetra, India, in 2006, both in Electronics and Communication Engineering, and the Ph.D. degree in Electrical Engineering from Indian Institute of Technology Delhi, India, in 2013. His PhD research was on crosstalk cancellation for vectored DSL systems. From  2012 to 2015, he was 
	with the CTO Office (Red Bank, NJ, USA), Ikanos Communications India Pvt. Ltd. (acquired by Qualcomm), Bangalore, India, where he was involved in research and development for xDSL systems.  Since 2015, he is  with the Faculty of Engineering, Bar-Ilan University, Ramat Gan, Israel as a post-doctoral researcher working on signal processing for wireline and wireless communications. He is an awardee of the  PBC (VATAT) Fellowship program for outstanding post-doctoral researchers from China and India from the Council for Higher Education, Israel  for two years (2016-2018).   His current research interests include signal processing for communications, communication protocols, multichannel communication, $5$G cellular networks, resource-limited  sensor networks,  and  xDSL systems.
	
	\item \textbf{Itsik Bergel} (itsik.bergel@biu.ac.il) received the B.Sc. in electrical  engineering (magna cum laude) and the B.Sc. in physics (magna cum laude) from Ben Gurion University, Beer-Sheva, Israel, in 1993 and 1994, respectively, and the M.Sc. (summa cum laude) and  Ph.D. in electrical engineering from the University  of Tel Aviv, Tel-Aviv, Israel, in 2000 and 2005  respectively. From 2001 to 2003, he was a Senior  Researcher at INTEL Communications Research Lab. In 2004 he received  the Yitzhak and Chaya Weinstein study award. In 2005, he was a Postdoctoral researcher at the  Dipartimento di Elettronica of Politecnico di Torino,  Italy, working on the capacity of non-coherent channels.  He is currently a faculty member
	in the faculty of engineering at Bar-Ilan University, Ramat-Gan, Israel. His main research interests include multichannel interference mitigation in wireline and wireless communications, cooperative transmission in cellular networks and cross layer optimization of random ad-hoc networks.
	Since
	2015, Dr. Bergel serves as an Associate Editor in the IEEE Transactions on Signal Processing. 
	\item \textbf{Amir Leshem} (leshema@biu.ac.il) received the B.Sc. (cum laude) in mathematics and physics, the M.Sc. (cum laude) in mathematics, and the Ph.D. degree in mathematics all from the Hebrew University, Jerusalem, Israel, in 1986, 1990, and 1998, respectively. From 1998 to 2000, he was with the Faculty of Information Technology and Systems, Delft University of Technology,
	The Netherlands, as a postdoctoral fellow working on algorithms for the reduction of terrestrial electromagnetic interference in radio-astronomical radio-telescope antenna arrays and signal processing for communication. From 2000 to 2003, he was Director of Advanced Technologies with Metalink Broadband where he was responsible for research
	and development of new DSL and wireless MIMO modem technologies and served as a member of several international standard setting groups such as ITU-T SG15, ETSI TM06, NIPP-NAI, IEEE 802.3  and 802.11. From 2000 to 2002, he was also a visiting researcher at Delft  University of Technology. In 2002 he joined Bar-Ilan University where he is one of the founders of the faculty of engineering and a full professor and head of the communications research track.   From 2003 to 2005, he was the technical manager of the  U-BROAD consortium developing technologies to provide 100 Mbps and  beyond over copper lines. In 2009 he spent his sabbatical at Delft University of Technology and Stanford University.  His main research interests include multichannel  wireless and wireline communication, applications of game theory to dynamic  and adaptive spectrum management of communication networks, array and  statistical signal processing with applications to multiple element sensor  arrays and networks, wireless communications, radio-astronomical imaging, set theory, logic and foundations of mathematics. Prof. Leshem was an Associate Editor of the IEEE Transactions on Signal Processing from 2008 to 2011, and he was the Leading Guest Editor  for special issues on signal processing for astronomy and cosmology in  IEEE Signal Processing Magazine and the IEEE Journal of Selected Topics in Signal Processing.

\end{itemize}

\end{document}